\documentclass[lettersize,journal]{IEEEtran}
\usepackage{amsthm, amssymb, amsmath, amsfonts}
\usepackage{algorithm}
\usepackage{algorithmic}
\usepackage{array}
\usepackage[caption=false,font=footnotesize,labelfont=rm,textfont=sf]{subfig}
\usepackage{textcomp}
\usepackage{stfloats}
\usepackage{url}
\usepackage{verbatim}
\usepackage{graphicx}
\usepackage{cite}
\usepackage{cuted}
\usepackage{multirow}
\usepackage{enumitem}
\usepackage[table,xcdraw]{xcolor}
\usepackage{diagbox}
\usepackage{tabularx}
\usepackage[colorlinks=true, linkcolor=blue, citecolor=blue, urlcolor=blue]{hyperref}
\usepackage[all]{hypcap}
\graphicspath{{figs/}}
\newcommand{\tabincell}[2]{\begin{tabular}{@{}#1@{}}#2\end{tabular}}

\renewcommand{\arraystretch}{1.5}

\newtheorem{lemma}{Lemma}
\newtheorem{corollary}{Corollary}
\newtheorem{remark}{Remark}
\newtheorem{proposition}{Proposition}

\begin{document}

\title{Revealing the Evanescent Components in Kronecker Product-based Codebooks: Insights and Applications}

\author{Jun Yang, Yijian Chen, Yunqi Sun, Yuan Si, Hongkang Yu, Shujuan Zhang and Zhaohua Lu
\thanks{The authors are with the State Key Laboratory of Mobile Network and Mobile Multimedia Technology, Shenzhen 518055, China, and the Wireless Product R\&D Institute, ZTE Corporation, Shenzhen 518055, China (emails: \{yang.jun10, chen.yijian, sun.yunqi, si.yuan, yu.hongkang, zhang.shujuan1, lu.zhaohua\}@zte.com.cn).}}



\maketitle

\begin{abstract}
The 2D DFT bases, derived from the Kronecker product of 1D DFT bases, form Kronecker product-based codebooks, which are used to construct Type I, Type II, and enhanced Type II codewords in 5G New Radio. Theoretically, each codeword in these codebooks should correspond to a beam directed towards a specific direction. However, this paper identifies a significant fraction of codewords in these codebooks as being associated with evanescent waves, rendering them redundant in array beamforming and far-field channel representation. This redundancy is analyzed using mathematical and electromagnetic models and verified through beam-pattern and system-level simulations. The spatial frequencies of these codewords are investigated based on spatial sampling theory, and codebooks for irregular arrays are proposed. Leveraging this redundancy, we propose a method to compress these codebooks, typically reducing their size by 22\%. The compression significantly decreases signaling and pilot overhead, and improves the efficiency of channel state information feedback and beam training without adding algorithmic complexity. The discussion is further extended to near-field and Rayleigh channel models, specifically regarding evanescent components, where guidelines for near-field codebook design and a new generation method for Rayleigh channels are provided.
\end{abstract}


\section{Introduction}\label{sec-1}
\IEEEPARstart{P}{recoding} is a critical technique for shaping the wavefront of wireless signals transmitted by phased arrays, thereby enhancing multiple-input multiple-output (MIMO) communications with improved spatial multiplexing, enhanced coverage and mitigated multi-user (MU) interference \cite{larsson14}. Codebook-based precoding and channel state information (CSI) feedback have been adopted in 4G and 5G communications. During channel estimation, the CSI is quantized by a codeword selected from a codebook. Kronecker product-based codebooks, designed for uniform planar arrays (UPAs), have been extensively discussed in MIMO communications \cite{lij13, wangy15, suhj17, huangy20}. Codewords in these codebooks serve as spatial-domain (SD) bases of Type I, Type II, and eType II codebooks in 5G New Radio (NR) \cite{38214r16}. A comprehensive overview of the evolution of NR codebooks from Release 8 to Release 17 can be found in \cite{qinz23, fux23}.

Type II codebooks, which have been part of the NR standard since Release 15, offer a more flexible representation of channel characteristics, making them particularly well-suited for MU-MIMO. The evolution of Type II codebooks continued with the introduction of enhanced Type II (eType II) and further enhanced Type II (feType II) codebooks, standardized in NR Releases 16 and 17, providing sufficient reduction on feedback overhead. Despite their potential advantages in MU MIMO, Type II codebooks and their evolved versions are still not supported by the vast majority of commercial user equipment (UE) due to the high implementation complexity and the significant feedback overhead. To fully realize the benefits of Type II codebooks, further efforts are required to tackle the implementation complexity and feedback overhead.

Codebook Subset Restriction (CBSR) has been introduced in the NR standard to directly reduce the codebook size \cite{38214r16}. In this approach, downlink signaling is transmitted prior to channel estimation to determine the set of SD bases available for CSI reporting. However, the primary purpose of CBSR is to mitigate MU interference rather than to reduce feedback overhead. Furthermore, in the case of large-scale arrays, the signaling overhead of CBSR itself becomes a significant concern due to the limited downlink signaling resources.

Future wireless communications are increasingly focused on deploying large-scale MIMO arrays, typically comprising hundreds to thousands of  antennas, to further improve spectral efficiency. Techniques such as subarray hybrid precoding \cite{luz24} and fast beam training \cite{wangx22, cuim23a, xuy24} have been proposed to accelerate the channel estimation for extremely large-scale aperture arrays (ELAAs). However, these methods introduce significant algorithmic complexity. For codebook-based feedback, achieving efficient CSI reporting is particularly challenging under stringent pilot and feedback overhead constraints. For instance, a Kronecker product-based codebook for a 64$\times$64 antenna array contains 65,536 codewords when oversampling factors of 4 are applied in both the horizontal and vertical dimensions. As a result, codebook-based beam training faces challenges in efficiently selecting a beam from a vast pool of candidate beams. Reducing the codebook size can help alleviate the difficulty of beam training for ELAAs.

\IEEEpubidadjcol

Conventional understanding holds that each codeword in a Kronecker product-based codebook corresponds to a beam within the half-space in front of the array to which it applies. This perception stems from the fact that 1D DFT bases represent a set of orthogonal beams covering an angle range of $[0,\pi]$ in a single dimension, and the Kronecker products of these bases simply populate the beams into a two-dimensional space that cover an angle range of $\mathcal{A}=[0,\pi]\times[0,\pi]$. However, the actual channel space is indeed a subset of $\mathcal{A}$, and the disregard of a crucial physical constraint during the Kronecker product operation on DFT bases should account for this discrepancy. Recent studies on electromagnetic (EM) based channel models and holographic MIMO have also revealed this physical constraint from spatial-frequency domain, i.e., physical channels are low-pass spatial filters \cite{pizzo20b, pizzo22, weil22, gongtr24}. However, such constraint has not yet been discussed from the view point of codebook design.

Since the space $\mathcal{A}$, spanned by the Kronecker product of DFT bases, is larger than the physical channel space, certain codewords in the codebooks may not correspond to any physical channel. Intriguingly, we found that these codewords represent EM components with higher wavenumbers, such as evanescent waves \cite{kong86}. Evanescent waves are special EM modes that exist in the reactive near-field region but contribute little to far-field transmission. While evanescent EM components may be used to enhance the capacity of a MIMO system \cite{jiran23}, their applications are typically limited to the reactive near-field region. Therefore, these codewords are unsuitable for far-field communications, and we refer to them as \textit{evanescent codewords}.

This paper aims to elucidate the presence of evanescent codewords in Kronecker product-based codebooks and proposes an optimization to these codebooks. The main contributions of this paper are summarized as follows.

\begin{itemize}
\item For the first time, we identify a critical deficiency in NR codebooks associated with evanescent codewords. This issue introduces unnecessary overhead and complexities in MIMO systems, notably impacting future standard designs. Based on spatial frequency theory, we also propose codebooks for irregular arrays.
\item We establish a connection between evanescent codewords and evanescent waves. We further reveal that these codewords are redundant and the low radiation efficiency of evanescent codewords is not due to spatial undersampling, highlighting a fundamental limitation in utilizing these codewords in array beamforming.
\item We propose compressing Kronecker product-based codebooks by excluding evanescent codewords. This compression enhances the efficiency of codebook-based CSI feedback and beam training without introducing algorithmic complexity. Notably, this compression does not compromise throughput performance.
\item We demonstrate that Fresnel near-field channels do not support evanescent components, confirming the effectiveness of the proposed codebook compression for both far-field and near-field MIMO systems. Additionally, we propose a method for generating Rayleigh channels that eliminates artifactual evanescent components.
\end{itemize}

The remainder of this paper is organized as follows: In Section \ref{sec-2}, we review the plane wave model and introduce evanescent EM components from the perspective of wavenumber. The line-of-sight (LOS) far-field channel model and Kronecker product-based codebooks are then briefly presented as preliminaries. A comprehensive investigation on evanescent codewords are present in Section \ref{sec-3}, focusing on elucidating their redundancy. Based on the analysis in Section \ref{sec-3}, Section \ref{sec-4} proposes an optimization for Kronecker product-based codebooks, accompanied by performance analysis and system-level simulations. In Section \ref{sec-5}, we extend the discussion to near-field and Rayleigh channel models to examine the presence of evanescent components in these channel models. Finally, conclusions are provided in Section \ref{sec-6}.

Notations: Vectors and matrices are denoted with boldfaced letters in this paper. The Cartesian coordinates system and spherical coordinates system as shown in Fig. \ref{f-xyz}a will be used interchangeably for array configurations and beam directions. For the Cartesian coordinates, it is in the form of $(x,y,z)$ while the coordinates in the spherical coordinate system are organized as $(r,\theta,\varphi)$, where $r$ is the radius, while $\theta$ and $\varphi$ are the elevation angle and azimuth, respectively. The direction of a beam is denoted by the elevation angle and azimuth $(\theta,\varphi)$.
\vspace{-0.5em}

\section{Preliminaries}\label{sec-2}
\subsection{The Plane-wave Model}
For the far-field propagation, a signal, typically referring to the electric field, transmitted or received by a MIMO array is commonly modeled as the superposition of a series of plane waves. A plane wave can be succinctly expressed as a vector field distribution in both temporal and spatial domains,
\begin{equation}
\mathbf{E}(\mathbf{r},t)=\mathbf{E}_0e^{j(\boldsymbol{k}\cdot\mathbf{r}-\omega t+\phi)},\label{eq-plane-wave1}
\end{equation}
where $\boldsymbol{k}$, $\mathbf{r}$ and $\phi$ denote the wave vector, the location vector and the initial phase, respectively. The wave vector specifies the direction of the wave propagation and its modulus $k=|\boldsymbol{k}|$ is the wavenumber. The angular frequency $\omega$ signifies the rate at which the wave oscillates over time at a particular location while the wavenumber $k$ portrays the pace at which the wave varies in space. Hence, $k$ and $\omega$ symbolize the spatial and temporal frequency of the plane wave respectively.

In free space, $k$ and $\omega$ are related to the carrier wavelength $\lambda$ by $k=\frac{2\pi}{\lambda}$ and $\omega=\frac{2\pi c}{\lambda}$ respectively, leading to $k=\frac{\omega}{c}$, where $c$ is the speed of light. Apparently, the temporal and spatial frequencies, while seemingly distinct, fundamentally depict the wave propagation from different perspectives, akin to the two sides of the same coin. The spatial frequency is typically defined as $\frac{1}{\lambda}$ in optics, whereas the wavenumber is commonly used in the context of radio waves. In this paper, the terms \textit{wavenumber} and \textit{spatial frequency} are used interchangeably to represent the same physical quantity, with $k$ specifically referring to the free-space wavenumber of a subcarrier.

\begin{figure}[h]
\centering
\subfloat[]{
\includegraphics[width=0.21 \textwidth]{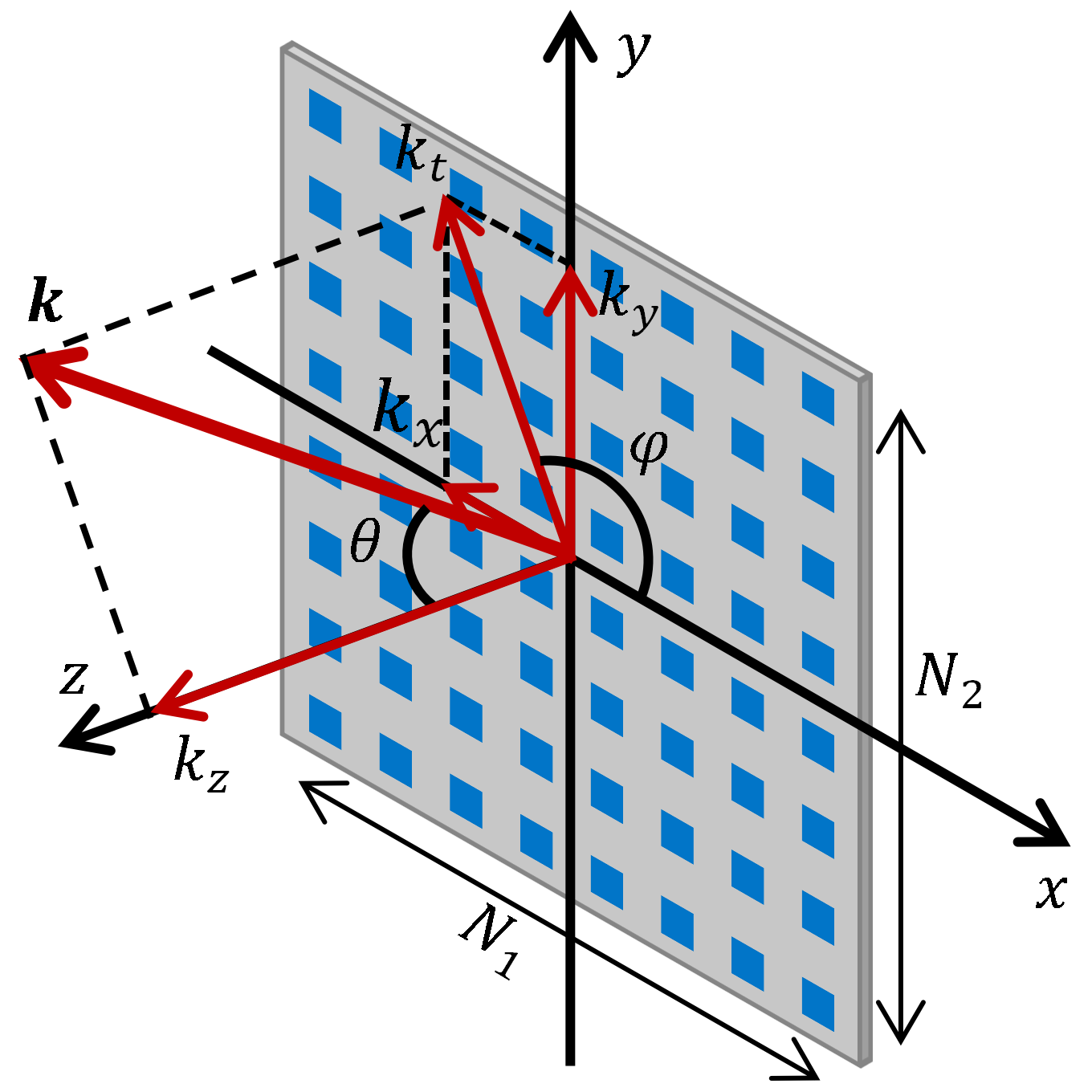}}\;
\subfloat[]{
\includegraphics[width=0.25 \textwidth]{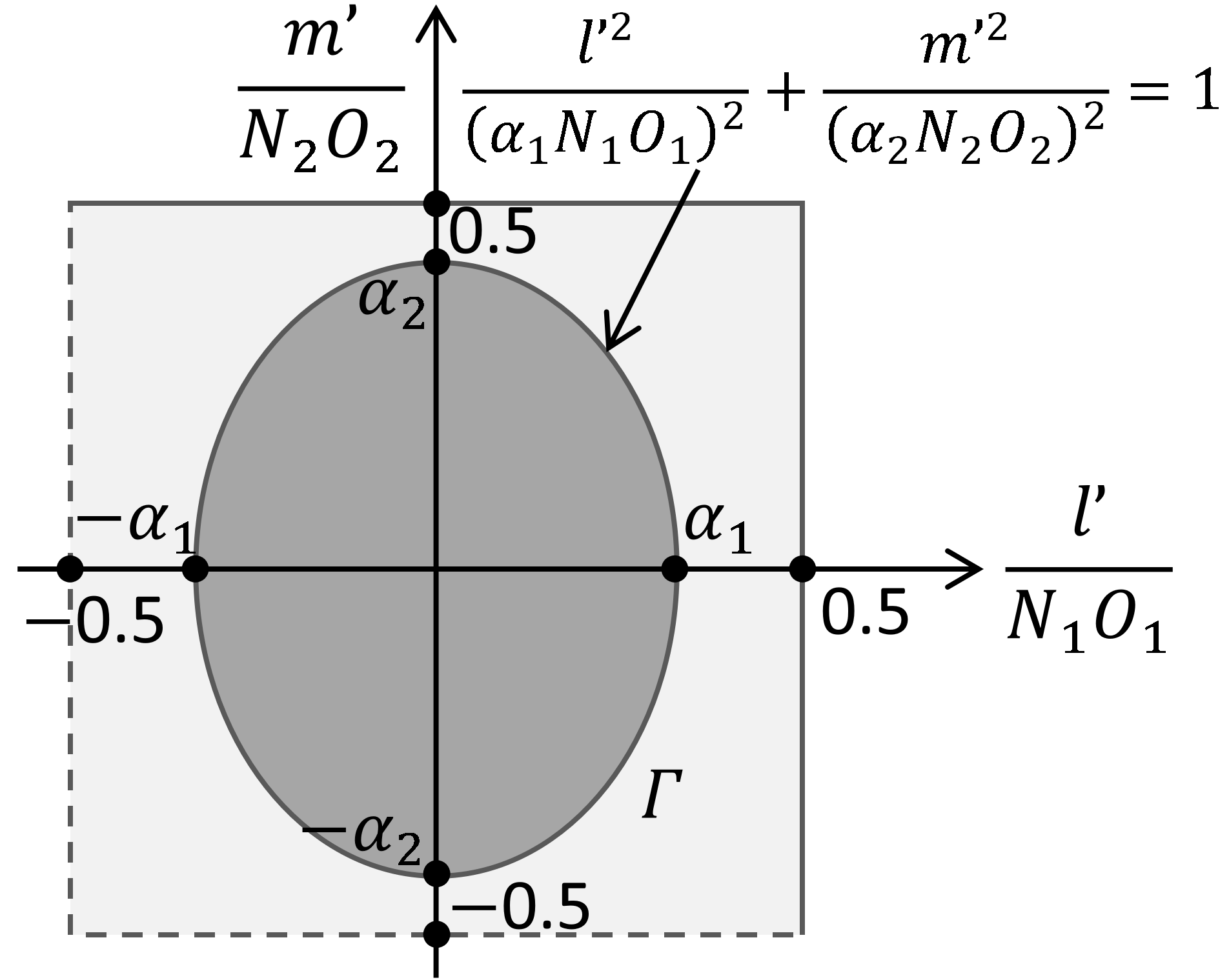}}
\caption{(a) The wave vector $\boldsymbol{k}$ in the reference Cartesian and spherical coordinate systems. (b) Illustration of the square confined by the indices of the codewords and the ellipse confined by inequality \eqref{eq-hf-area}.}
\label{f-xyz}
\end{figure}

In the Cartesian coordinate system as shown in Fig. \ref{f-xyz}a, let $\boldsymbol{k}=(k_x, k_y,k_z)$ and $\mathbf{r}=(x,y,z)$, the plane wave model \eqref{eq-plane-wave1} can be rewritten as
\begin{equation}
\mathbf{E}(\mathbf{r},t)=\mathbf{E}_0(t)e^{j(k_{x}x+k_{y}y+k_{z}z)},\label{eq-plane-wave2}
\end{equation}
where $\mathbf{E}_0(t)=\mathbf{E}_0e^{j(\omega t+\phi)}$. When the spatial characteristics is of concern, the time-varying part $\mathbf{E}_0(t)$ is regarded as a constant complex amplitude.

\begin{remark}\label{rmk-1}
The phase of a plane wave across any sampling plane varies linearly, with a constant phase gradient equal to the projection of $\boldsymbol{k}$ onto the plane.
\end{remark}

As illustrated in Fig. \ref{f-xyz}a, if a plane wave propagates with wave vector $\boldsymbol{k}$ in free space, the phase distribution in the $x$-$o$-$y$ plane exhibits a gradient $k_t = |\boldsymbol{k}|\sin\theta$, where $\theta$ denotes the angle between $\boldsymbol{k}$ and the plane's normal direction, i.e., the elevation angle. Notably, $k_t$ determines the propagation mode of the plane wave relative to the sampling plane and remains bounded by $k$, irrespective of the plane's orientation. The following dispersion relation holds for the wave vector of a monochromatic plane wave\cite{kong86}
\begin{equation}
k_x^2+k_y^2+k_z^2=k^2. \label{eq-kxkykz}
\end{equation}
If the dispersion relation is violated during wave propagation, it implies that the EM wave has either been modulated to another frequency or has entered a medium with a different refractive index.

In classic EM theory, there exists a special class of solutions to the Maxwell's equations in which $k_x^2+k_y^2>k^2$\cite{kong86}. If the dispersion relation \eqref{eq-kxkykz} is to hold true, $k_z$ should become an imaginary wavenumber, which leads to an exponential decay to the wave amplitude when it propagates towards the $z$ direction. Such solutions are referred to as evanescent waves, and a typical case that an evanescent wave occurs is the total internal reflection on a glass-air interface\cite{bennett08}. Due to the exponential decay, the wave cannot propagate effectively in the $z$ direction.

On the other hand, the wave can still propagate in the $x$-$o$-$y$ plane, but with a higher transverse spatial frequency. An evanescent wave belongs to a non-propagation mode, typically arising on sub-wave-length structures or interfaces where discontinuity or inhomogeneity exists\cite{harrington60, fink07, milosevic13}. Hence, the evanescent wave usually remains confined near an interface (``reactive boundary'' in \cite{harrington60}) and the propagation along the interface exhibits primarily oscillatory behavior. Since the evanescent wave has a higher transverse spatial frequency, it possesses a shorter wavelength compared to a plane wave with the same temporal frequency, and it propagates slower along the surface in which it is guided. There are also complicated transition mechanisms between the evanescent wave and the propagating wave \cite{pendry04, fink07} on sub-wavelength structures. In general, an evanescent wave component is not able to propagate to far field directly, rendering it impractical for long-distance communications.

\subsection{The Far-field Channel Model}\label{sec-ff-channel}
The far-field transmission can be regarded as a match between the signal propagation mode and the wireless channel. The propagation mode of a plane wave is characterized by the phase gradient of the wave on the array plane as it determines the beam direction. As shown in Fig. \ref{f-xyz}a, for a UPA with $N_1$ antennas in the $x$-direction and $N_2$ antennas in the $y$-direction, the lossless LOS channel between the antenna located at the $n_1$-th row and $n_2$-th column of the array and a far-field receiving antenna in the $(\theta,\varphi)$ direction can be expressed as
\begin{equation}
h_{n_1n_2}(\theta,\varphi)=e^{-j\left(k_xn_1d_1+k_yn_2d_2\right)},\label{eq-h}
\end{equation}
where
\begin{equation}
k_x=k\sin\theta\cos\varphi,\quad k_y=k\sin\theta\sin\varphi,
\end{equation}
and $n_1\in\{0,1,\cdots,N_1-1\}$, $n_2\in\{0,1,\cdots,N_2-1\}$. The antenna spacing in the $x$- and $y$-directions are denoted by $d_1$ and $d_2$, respectively.

\begin{remark}\label{rmk-2}
By comparing equation \eqref{eq-h} and \eqref{eq-plane-wave2}, we observe that the propagation of a plane wave is governed by the wavenumbers $k_x$, $k_y$ and $k_z$, while the wireless channel associated with this EM wave specifies only $k_x$ and $k_y$. Although $k_z$ is not explicitly expressed in the channel model, it is determined by $k_x$ and $k_y$ via the dispersion relation \eqref{eq-kxkykz}. 
\end{remark}

Hence, when an array lying in the $x$-$o$-$y$ plane is to generate a beam along the direction indicated by $\boldsymbol{k}=(k_x, k_y,k_z)$, the precoding is determined solely by $\boldsymbol{k}_t=(k_x,k_y)$. For any valid propagation mode with respect to the array plane, $k_x$, $k_y$ and $k_z$ must be real, and the transverse wavenumber $k_t=(k_x^2+ k_y^2)^{\frac{1}{2}}$ cannot exceed $k$. In this sense, the LOS wireless channel is a low-pass spatial filter, as demonstrated in \cite{pizzo20b} and many other studies. Consequently, evanescent components are inherently filtered out by the far-field channel.
\vspace{-0.8em}

\subsection{Kronecker product-based Codebooks}

For the UPA shown in Fig. \ref{f-xyz}a, the precoding for directing a beam towards a particular direction can be achieved using the Kronecker product of two steering vectors that represent the phase gradients $k_x$ and $k_y$, respectively. In the NR standard, a codeword for a UPA can be constructed by the Kronecker product of two 1D DFT bases \cite{38214r16},
\begin{equation}
\begin{aligned}
\boldsymbol{v}_{lm}=&\left\{\left[1,e^{j\frac{2\pi l}{N_1O_1}},\cdots,e^{j(N_1-1)\frac{2\pi l}{N_1O_1}}\right]\right.\otimes\\
&\;\;\left.\left[1,e^{j\frac{2\pi m}{N_2O_1}},\cdots,e^{j(N_2-1)\frac{2\pi l}{N_2O_2}}\right]\right\}^{T},
\end{aligned}\label{eq-vlm}
\end{equation}
where $l\in\{0,1,\cdots,N_1O_1-1\}$, $m\in\{0,1,\cdots,N_2O_2-1\}$, with $O_1$ and $O_2$ denoting the oversampling factors. There are $N_1O_1$ and $N_2O_2$ 1D DFT bases in the horizontal and vertical directions respectively and the codebook comprises a total of $N_1O_1N_2O_2$ codewords.

The indices $l$ and $m$ in $\boldsymbol{v}_{lm}$ determines the phase gradients of the codeword in two orthogonal directions,
\begin{equation}
k_1=\frac{2\pi l}{N_1O_1},\quad k_2=\frac{2\pi m}{N_2O_2}.\label{eq-nominal-k}
\end{equation}
We need to emphasize that $k_1$ and $k_2$ are the \textit{nominal phase gradients}. The actual phase gradients of the codeword depend on the antenna spacing of the array on which the codeword is applied. A same codeword will exhibit a different spatial phase gradient on arrays with varying antenna spacing, hence generating beams at different directions.

This paper focuses on the analysis for Kronecker product-based codebooks, thus all subsequent references to codebooks and codewords will pertain to those based on the Kronecker product, unless otherwise specified. For simplicity, a Kronecker product-based codebook will be denoted as $\mathbf{CB}_{N_1, N_2}^{O_1, O_2}$.

\section{The Evanescent Codewords}\label{sec-3}
While the Kronecker product operation enables structured indexing in the codebooks, it also introduces redundancy. In this section, we analyze this redundancy from both mathematical and electromagnetic perspectives and offer a comprehensive discussion of its implications and related aspects.
\vspace{-0.6em}
\subsection{The Mathematical Aspect}
Theoretically, a codeword $\boldsymbol{v}_{lm}$ should represent a beam that matches a particular LOS channel, i.e., $v_{lm,n_1n_2}=h_{n_{1}n_{2}}^{*}(\theta,\varphi)$, which yields
\begin{equation}
e^{j2\pi\left(\frac{n_1l}{N_1O_1}+\frac{n_2m}{N_2O_2}\right)}=e^{j(n_1k_xd_1+n_2k_yd_2)}.\label{eq-channel-match}
\end{equation}
To simplify the analysis, we rewrite \eqref{eq-channel-match} as
\begin{equation}
e^{j2\pi\left(\frac{n_1l'}{N_1O_1}+\frac{n_2m'}{N_2O_2}\right)}=e^{j(n_1k_xd_1+n_2k_yd_2)},\label{eq-channel-match2}
\end{equation}
where
\begin{equation}
l'=\left\{
\begin{aligned}
&l,&&l\leq N_1O_1/2,\\
&l-N_1O_1, &&\text{otherwise},
\end{aligned}\right.\label{eq-l-prime}
\end{equation}
\begin{equation}
\quad m'=\left\{
\begin{aligned}
&m,&&m\leq N_2O_2/2,\\
&m-N_2O_2, &&\text{otherwise}.
\end{aligned}\right.\label{eq-m-prime}
\end{equation}
The substitution of $(l,m)$ by $(l',m')$ shifts the ranges of the nominal phase gradients $k_1$ and $k_2$ in \eqref{eq-nominal-k} from $[0, 2\pi)$ to $(-\pi, \pi]$, leveraging the identity $e^{ja} = e^{j(a + 2n\pi)}$, where $a \in \mathbb{R}$ and $n \in \mathbb{Z}$. Since $k_x=k\sin\theta\cos\varphi$ and $k_y=k\sin\theta\sin\varphi$, we obtain
\begin{equation}
\begin{aligned}
&2\pi(n_1\alpha_1\sin\theta\cos\varphi+n_2\alpha_2\sin\theta\sin\varphi)\\
&\qquad=2\pi\left(\frac{n_1l'}{N_1O_1}+\frac{n_2m'}{N_2O_2}\right)+2A\pi,\; A\in\mathbb{Z},
\end{aligned}\label{eq-phase-match}
\end{equation}
where $\alpha_1=\frac{d_1}{\lambda}$ and $\alpha_2=\frac{d_2}{\lambda}$ are the  \textit{normalized antenna spacing}. Equation \eqref{eq-phase-match} should hold for UPAs with different number of antennas, i.e. $n_1$ and $n_2$ can be any integers, hence it can be decoupled into the following two equations,
\begin{align}
&\sin\theta\cos\varphi=\frac{1}{\alpha_1}\left(\frac{l'}{N_1O_1}+A_1\right),\label{eq-lm-restrict1}\\
&\sin\theta\sin\varphi=\frac{1}{\alpha_2}\left(\frac{m'}{N_2O_2}+A_2\right),\label{eq-lm-restrict2}
\end{align}
where $A_1,A_2\in\mathbb{Z}$. The beam direction $(\theta,\varphi)$ can then be obtained by solving \eqref{eq-lm-restrict1} and \eqref{eq-lm-restrict2}.

\begin{proposition}\label{prop-1}
A codeword $\boldsymbol{v}_{lm}$ defined by \eqref{eq-vlm} is incapable of generating a directional beam on a UPA with normalized antenna spacings $\alpha_1$ and $\alpha_2$ if its index $(l,m)$ does not satisfy the following inequality,
\begin{equation}
\left(\frac{l'}{\alpha_1N_1O_1}\right)^2+\left(\frac{m'}{\alpha_2N_2O_2}\right)^2\leq 1,\label{eq-hf-area}
\end{equation}
where $(l,m)$ relates to $(l',m')$ by equations \eqref{eq-l-prime} and \eqref{eq-m-prime}.
\end{proposition}
\begin{IEEEproof}
Squaring both sides of equations \eqref{eq-lm-restrict1} and \eqref{eq-lm-restrict2}, and summing the results yields
\begin{equation}
\sin^{2}\theta=\frac{1}{\alpha_1^2}\left(\frac{l'}{N_1O_1}+A_1\right)^2+\frac{1}{\alpha_2^2}\left(\frac{m'}{N_2O_2}+A_2\right)^2.\label{eq-sin2}
\end{equation}
Given that $\sin^2\theta\in[0,1]$ for any real $\theta$, it necessitates the following inequality,
\begin{equation}
\frac{1}{\alpha_1^2}\left(\frac{l'}{N_1O_1}+A_1\right)^2+\frac{1}{\alpha_2^2}\left(\frac{m'}{N_2O_2}+A_2\right)^2\leq 1.\label{eq-p1-constraint}
\end{equation}
To ensure this inequality is satisfied, $A_1$ and $A_2$ should be selected carefully to minimize $|\frac{l'}{N_1O_1}+A_1|$ and $|\frac{m'}{N_2O_2}+A_2|$. Since $\frac{l'}{N_1O_1}$ and $\frac{m'}{N_2O_2}$ are both confined in the range $(-0.5,0.5]$, it follows that $|\frac{l'}{N_1O_1}+A_1|\geq |\frac{l'}{N_1O_1}|$ and $|\frac{m'}{N_2O_2}+A_2|\geq |\frac{m'}{N_2O_2}|$ for $A_1,A_2\in \mathbb{Z}$. Hence, $A_1$ and $A_2$ must be zero, simplifying the inequality \eqref{eq-p1-constraint} to \eqref{eq-hf-area}.  If $\boldsymbol{v}_{lm}$ does not satisfy \eqref{eq-hf-area}, it leads to a beam direction with a complex elevation angle according to \eqref{eq-sin2}, indicating a beam in a non-physical space. As a result, the codeword cannot fulfill its primary function, i.e., generating a directional beam in physical space.
\end{IEEEproof}
Proposition \ref{prop-1} can also be elucidated from a geometric perspective. The coordinate $(\frac{l'}{N_1O_1},\frac{m'}{N_2O_2})$ is confined in a square defined by (-0.5,0.5]$\times$(-0.5,0.5] while inequality \eqref{eq-hf-area} delineates an ellipse with semi-major axis $\alpha_1$ and semi-minor axis $\alpha_2$, as illustrated in Fig. \ref{f-xyz}b. To satisfy \eqref{eq-p1-constraint}, $(\frac{l'}{N_1O_1},\frac{m'}{N_2O_2})$ must reside within the intersection of the square and the ellipse. If $A_1$ or $A_2$ is non-zero, the square will shift away from the ellipse, resulting in less or no overlapping.

Based on equations \eqref{eq-lm-restrict1} and \eqref{eq-lm-restrict2}, the azimuth of a beam corresponding to codeword $\boldsymbol{v}_{lm}$ can be obtained as following,
\begin{equation}
\varphi=\arctan\left(\frac{m'}{l'}\frac{\alpha_1N_1O_1}{\alpha_2N_2O_2}\right).\label{eq-varphi}
\end{equation}
The azimuth $\varphi$ is real for all codewords and it indicates the direction of the transverse wave vector $\boldsymbol{k}_t$ of the wave component represented by $\boldsymbol{v}_{lm}$.

\begin{figure}[b]
\centering{\includegraphics[width=0.48 \textwidth]{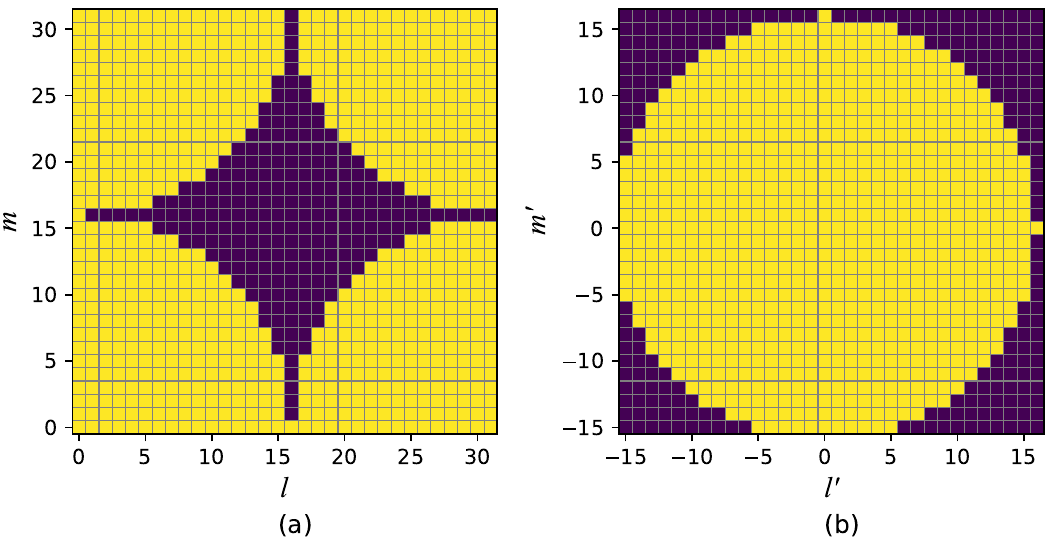}}
\caption{The distribution of propagating and evanescent codewords in the index space for $N_1=N_2=8$, $O_1=O_2=4$ and $\alpha_1=\alpha_2=0.5$. The dark purple patches indicate the evanescent codewords.}
\label{f-hf-area}
\end{figure}

In this paper, a $\boldsymbol{v}_{lm}$ is termed an \textit{evanescent codeword} if its index $(l,m)$ violates \eqref{eq-hf-area}; otherwise, it is considered as a \textit{propagating codeword}. Fig. \ref{f-hf-area} shows the distribution of evanescent codewords in $\mathbf{CB}_{8,8}^{4,4}$ for $\alpha_1=\alpha_2=0.5$ in the index space. The two subfigures indicates that the substitution of $(l,m)$ by $(l', m')$ is equivalent to a DFT-shift that relocates the zero-frequency component to the center. The evanescent codewords gather in a specific region in the codebooks, as shown in Fig. \ref{f-hf-area}, termed the \textit{evanescent zone}, while the remaining region is referred to as the \textit{propagating zone}. Fig. \ref{f-hf-area}b is indeed a pixelated version of Fig. \ref{f-xyz}b, and the jagged boundary between the evanescent and propagating zones approaches a perfect ellipse when $O_1,O_2\to \infty$. The zone boundary $\Gamma$ corresponds to $\theta=90^{\circ}$, i.e., the equality in \eqref{eq-hf-area} is satisfied.

\begin{corollary}\label{coro-1}
The number of evanescent codewords, $N_{\text{eva}}$, in a codebook depends on the normalized antenna spacings, $\alpha_1$ and $\alpha_2$, of the array to which it applies. If $\alpha_1=\alpha_2\geq\frac{\sqrt{2}}{2}$, the codebook contains no evanescent codewords.
\end{corollary}
\begin{IEEEproof}
This dependency of $N_{\text{eva}}$ on array antenna spacing follows directly from the inequality \eqref{eq-hf-area} and is illustrated in Fig. \ref{f-xyz}b. When $\alpha_1=\alpha_2>\frac{\sqrt{2}}{2}$, the inequality \eqref{eq-hf-area} is satisfied for all codewords, irrespective of $N_1$, $N_2$, $O_1$ and $O_2$.
\end{IEEEproof}
The redundancy can be calculated as $r_{\text{eva}}=\frac{N_{\text{eva}}}{N_{\text{tot}}}\times 100\%$, where $N_{\text{tot}}$ denotes the total number of codewords. As illustrated in Fig. \ref{f-xyz}b, $r_{\text{eva}}=(1-\frac{A_o}{A_s})$ when $O_1, O_2\to\infty$, where $A_s$ denotes the area of the square while $A_o$ is the area of the overlapping region between the square and ellipse. Notably, $r_{\text{eva}}=(1-\frac{\pi}{4})\approx$ 22\% when $\alpha_1=\alpha_2=0.5$. Half-wavelength antennas are widely adopted in commercial MIMO arrays for their higher radiation efficiency and their ability to prevent grating lobes \cite{balanis15}. However, their use inevitably introduces evanescent codewords, making codeword redundancy a prevalent concern in MIMO communications.

\begin{corollary}\label{coro-2}
A 1D DFT codebook for a uniform linear array (ULA) contains no evanescent codewords if the antenna spacing of the ULA is no less than half the wavelength.
\end{corollary}
\begin{IEEEproof}
For a 1D DFT codebook, the index $m'=0$ and \eqref{eq-hf-area} is simplified to $\left|\frac{l'}{\alpha_1N_1O_1}\right|\leq 1$. Since $\frac{l'}{N_1O_1}\in[-0.5,0.5]$, this inequality is always satisfied if $\alpha_1\geq 0.5$.
\end{IEEEproof}
As previously mentioned, commercial MIMO arrays are commonly equipped with half-wavelength antennas. In such cases, evanescent codewords do not appear in 1D DFT codebooks. Therefore, this paper focuses on the analysis of 2D codebooks.
\vspace{-0.3em}

\subsection{The Electromagnetic Aspect}
According to \eqref{eq-channel-match2}, the index of a codeword relates to the transverse wave vector $\boldsymbol{k}_t=(k_x,k_y)$ by $k_x=\frac{2\pi}{d_1}\frac{l'}{N_1O_1}$ and $k_y=\frac{2\pi}{d_2}\frac{m'}{N_2O_2}$. Given that $k=\frac{2\pi}{\lambda}$, $\alpha_1=\frac{d_1}{\lambda}$ and $\alpha_2=\frac{d_2}{\lambda}$, it follows that
\begin{equation}
k_x=\frac{l'}{\alpha_1N_1O_1}k,\quad
k_y=\frac{m'}{\alpha_2N_2O_2}k.\label{eq-kxky}
\end{equation}
Substituting the above equations into \eqref{eq-hf-area} yields an equivalent condition for identifying an evanescent codeword based on its spatial frequency,
\begin{equation}
k_x^2+k_y^2\leq k^2. \label{eq-kx-ky}
\end{equation}
We can infer from \eqref{eq-kx-ky} that, if the index of a codeword fails to satisfy \eqref{eq-hf-area}, the resulting wave component has a transverse wavenumber that surpasses $k$. As demonstrated in Section \ref{sec-2}, the violation of \eqref{eq-kx-ky} leads to the formation of an evanescent wave. Consequently, the codewords within the evanescent zone correspond to evanescent components with transverse spatial frequencies higher than $k$.

According to \eqref{eq-kxky}, the index space $(l',m')$ can be mapped to the $(k_x,k_y)$ space by scaling with the stretch factors $\frac{k}{\alpha_1N_1O_1}$ and $\frac{k}{\alpha_2N_2O_2}$ in the $l'$- and $m'$-directions, respectively. The farther an evanescent codeword is away from the zone boundary $\Gamma$, the higher the spatial frequency it stands for. In the following discussions, referencing the spatial frequency of a codeword is equivalent to referencing the transverse spatial frequency of the EM component that the codeword represents.

The spatial frequencies $k_x$ and $k_y$ of a codeword are in fact the phase gradients in two orthogonal directions. Hence, the phase at any position $(x,y)$ on the array plane can be calculated as $\phi(x,y)=k_x x+k_y y$. Then, inspired by \eqref{eq-kxky}, the codeword for an irregular planar array with arbitrary antenna layout can be constructed as
\begin{equation}
\boldsymbol{u}_{l,m}^{2D}=e^{jk\left( \frac{l}{\alpha_1N_1O_1}\mathbf{X}+\frac{m}{\alpha_2N_2O_2}\mathbf{Y}\right)},
\end{equation}
where $\mathbf{X}$ and $\mathbf{Y}$ are vectors storing the $x$- and $y$-coordinates of the antennas. This codebook can be further extended to a more general one, which supports arbitrary 3D arrays, as
\begin{equation}
\boldsymbol{u}_{l,m}^{3D}=e^{jk\left( \frac{l}{\alpha_1N_1O_1}\mathbf{X}+\frac{m}{\alpha_2N_2O_2}\mathbf{Y}
+\tilde{k}_z\mathbf{Z}\right)},
\end{equation}
where $\tilde{k}_z=\sqrt{1-\left( \frac{l'}{\alpha_1N_1O_1}\right)^2-\left(\frac{m'}{\alpha_2N_2O_2}\right)^2}$ and the vector $\mathbf{Z}$ stores the $z$-coordinates of the antennas.

Clearly, codewords $\boldsymbol{u}_{l,m}^{2D}$ and $\boldsymbol{u}_{l,m}^{3D}$ are similar to Kronecker product-based codewords and they can still be indexed by $(l,m)$. Hence, the CSI feedback is compatible with the current NR standard. Furthermore, these codewords may represent evanescent waves if their indices violate \eqref{eq-hf-area}.
\vspace{-0.5em}

\subsection{Beam Pattern Analysis}\label{sec-fullwave-sim}
To verify Proposition \ref{prop-1}, a beamforming model similar to the one for reconfigurable intelligent surface \cite[Eqs. (6) and (7)]{juny23} is adopted. Assume that an array is equipped with $N$ antennas, where the $i$-th antenna locates at $\mathbf{r}_{i}$, $i\in{1,2, \cdots,N}$ and a single-antenna UE receives the transmitted signal at $\mathbf{r}'$, then the received power can be expressed as
\begin{equation}
P(\mathbf{r}')=S\left|\sum_{i=1}^{N}\frac{\sqrt{P_{i}F_n(\theta_{i},\varphi_{i})}}{|\mathbf{\mathbf{r}_{i}-\mathbf{r}'}|}e^{-j(k|\mathbf{\mathbf{r}_{i}-\mathbf{r}'}|+\phi_n)}\right|^2,\label{eq-pr}
\end{equation} 
where $P_{i}$ and $F_n(\theta_{i},\varphi_{i})$ denote the transmission power and power-normalized radiation pattern of the $i$-th antenna, respectively, $(\theta_{i},\varphi_{i})$ being the angles of departure with respect to the $i$-th antenna. The phase $\phi_n$ on the $n$-th antenna is the imposed precoding phase-shift. The effective aperture of the UE antenna $S$ is presumed to be the actual area of the antenna (e.g., $S=\lambda^2/4$). To simplify the analysis, isotropic radiation pattern is assumed for the transmission antenna, i.e. $F_n=\frac{1}{2\pi}$.

\begin{figure}[t]
\centering
\subfloat[]{
\includegraphics[width=0.23 \textwidth, trim=3.2cm 0.8cm 0.9cm 3.2cm, clip]{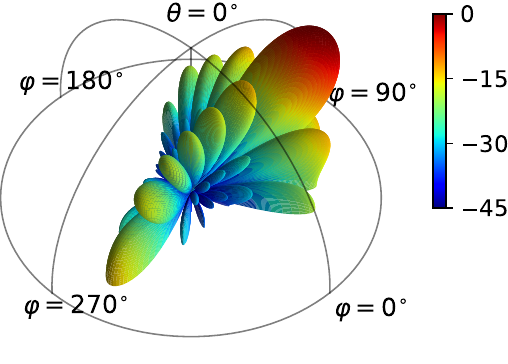}}\;
\subfloat[]{
\includegraphics[width=0.23 \textwidth, trim=3.2cm 0.8cm 0.9cm 3.2cm, clip]{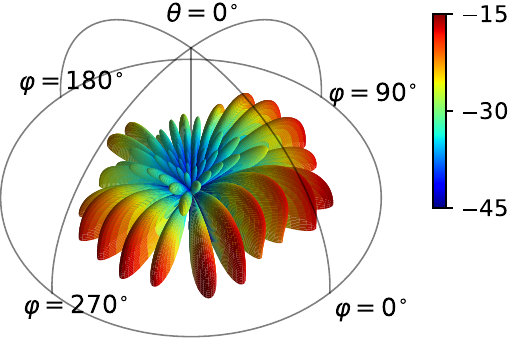}}\\
\subfloat[]{
\includegraphics[width=0.19 \textwidth]{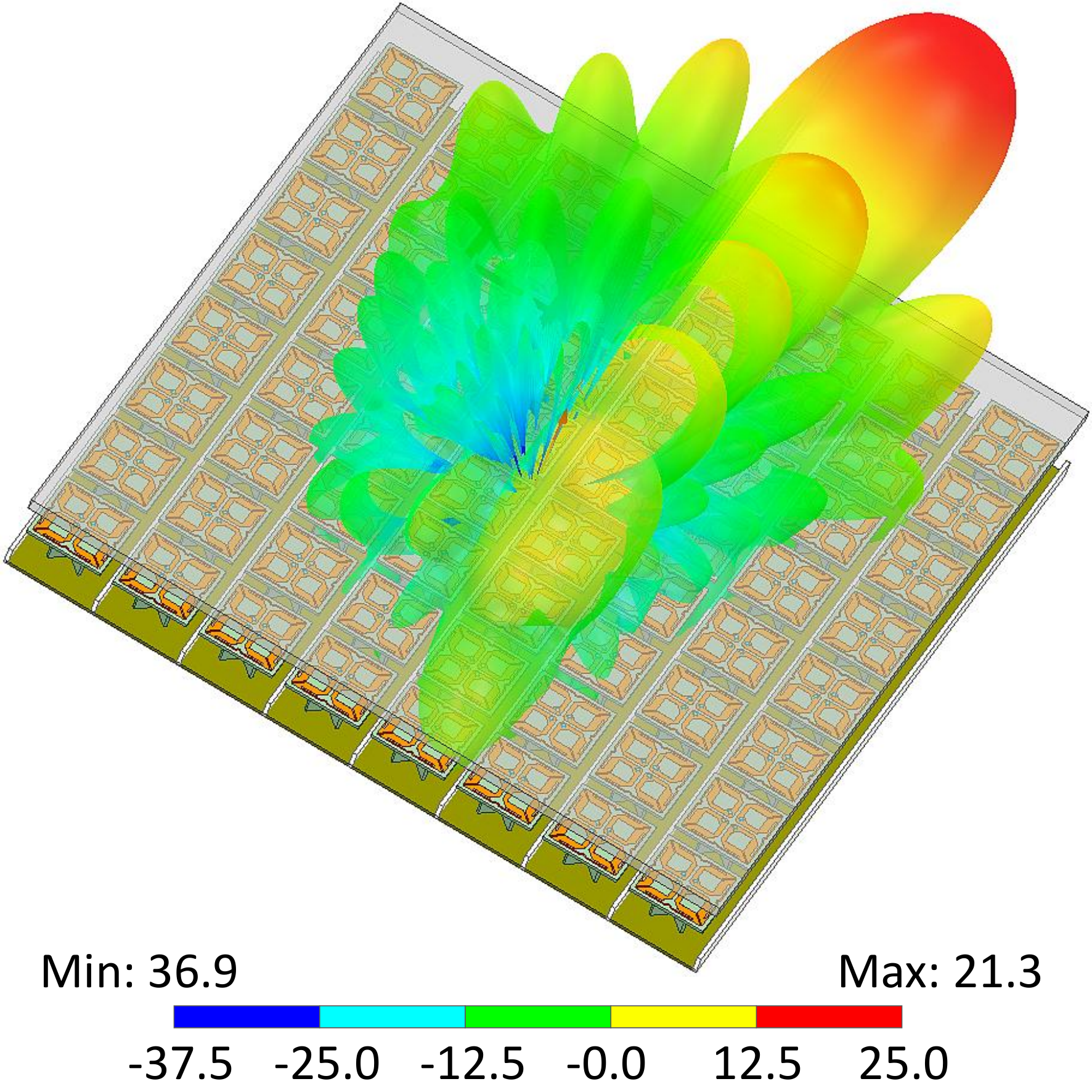}}\qquad\quad
\subfloat[]{
\includegraphics[width=0.19 \textwidth]{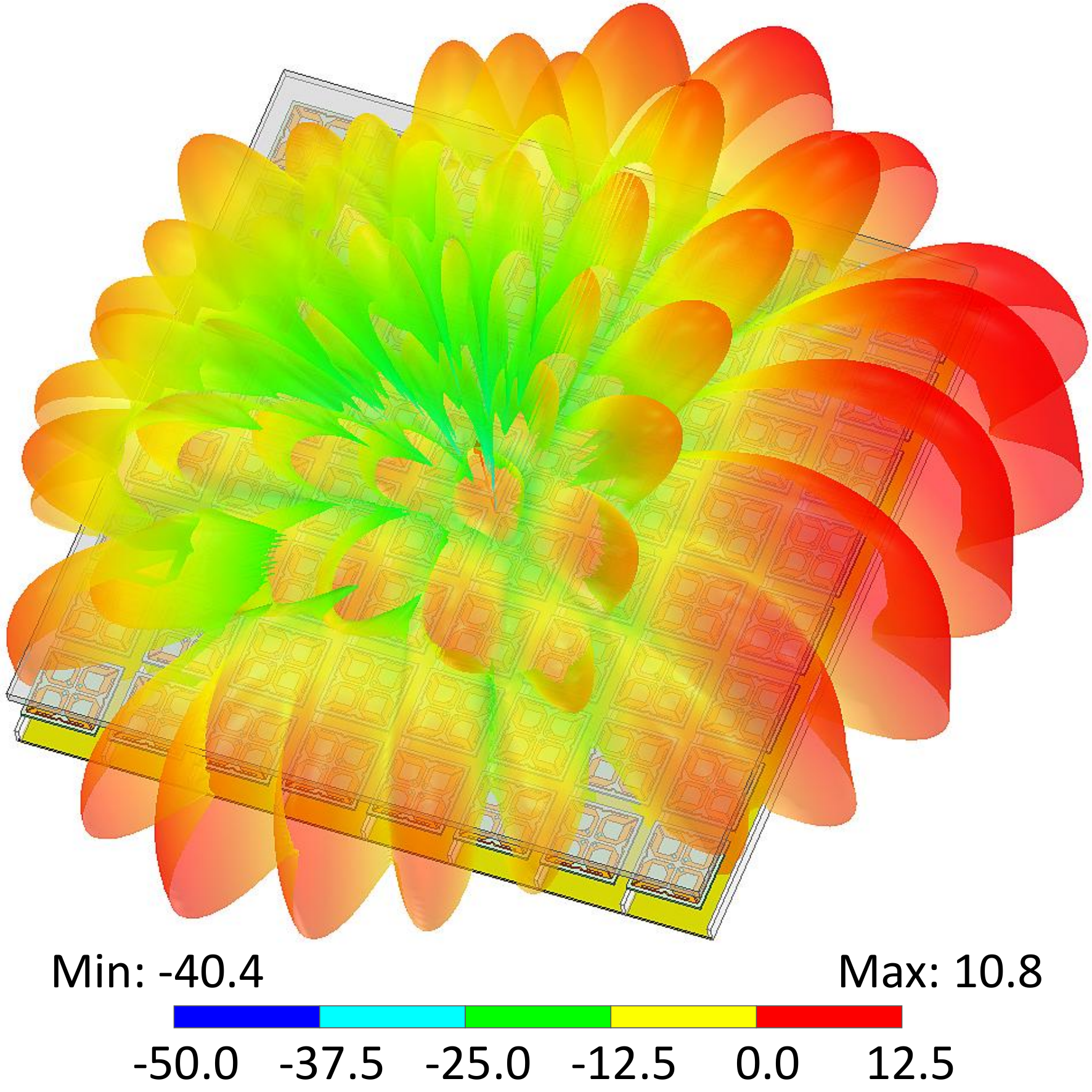}}
\caption{Beam patterns of codewords $\boldsymbol{v}_{4,10}$ (the left column) and $\boldsymbol{v}_{14,16}$ (the right column) from $\mathbf{CB}_{8,8}^{4,4}$. Subfigures (a) and (b) are based on array synthesis while (c) and (d) are results of full-waveform simulations.}
\label{f-beam-pat-sim}
\vspace{-0.8em}
\end{figure}

Figs. \ref{f-beam-pat-sim}a and \ref{f-beam-pat-sim}b show respectively the beam patterns of the propagating codeword $\boldsymbol{v}_{4,10}$ and the evanescent codeword $\boldsymbol{v}_{14,16}$ from $\mathbf{CB}_{8,8}^{4,4}$. A directional beam towards $\theta =42.3^{\circ}$ and $\varphi =68.2^{\circ}$ is generated by $\boldsymbol{v}_{4,10}$ while no distinguishable main lobe appears in the beam pattern of $\boldsymbol{v}_{14,16}$. When $\boldsymbol{v}_{14,16}$ is applied, the energy primarily radiates along the edge of the array while no significant radiation is observed in the normal direction, consistent with the aforementioned characteristics of evanescent waves. The beam pattern of $\boldsymbol{v}_{14,16}$ suggests that the energy is not completely filtered out. This ``leakage'' is due to the finite aperture of the UPA, which imposes a sinc-function type filter which allows a small portion of out-of-band energy to pass through.

In regard to practical arrays with dedicated antenna patch designs, complex mutual coupling exists among the antennas \cite{chenx18}, and the radiation pattern of the antennas deviates from the ideal isotropic one. Moreover, transitions between evanescent and propagating modes can occur via sub-wavelength structures in the array. The above array synthesis-based simulations with point source assumption falls short of capturing the mutual coupling and EM mode transition. Conversely, full-waveform methods, such as discontinuous Galerkin methods \cite{juny17}, is able to simulate the interaction between the radiated EM field and the antenna structure by solving Maxwell's equations with specific boundary conditions. Hence, full-waveform simulations were conducted to further confirm the beam pattern of evanescent codewords.

An antenna array with 8$\times$8 dual-polarized half-wavelength ($\approx$ 22.4$\times$22.4 mm$^2$) antennas, operating at the 6.7 GHz band, was employed for the full-waveform simulations. As depicted in Figs. \ref{f-beam-pat-sim}c and \ref{f-beam-pat-sim}d, the full-waveform beam patterns are similar to the results from the array synthesis, which indicates that the mutual coupling does not significantly affect the radiation pattern. According to the full waveform simulations, the maximum radiation power under the precoding $\boldsymbol{v}_{14,16}$ is about 10.5 dB lower than that of $\boldsymbol{v}_{4,10}$. Additionally, stronger radiation in the normal direction is observed in the full-waveform simulation, which may indicate that a small portion of evanescent components is transformed into propagating components. Both array synthesis-based and full-waveform simulations show that evanescent codewords cannot generate directional beams in physical space.

The comparison also indicates that array synthesis-based simulations are sufficiently accurate for beam pattern analysis when mutual coupling is negligible. In the subsequent sections, the array synthesis-based method is adopted, as it is far more efficient than full-waveform simulations.
\vspace{-0.5em}

\subsection{Spatial Sampling Criterion for Evanescent Codewords}
The precoding of a MIMO array can be considered as a spatial sampling of a particular EM wave, and it reconstructs the wave propagation provided that the spatial sampling satisfies the Nyquist criterion, otherwise it gives rise to grating lobes. Evanescent codewords represent EM components with higher spatial frequencies and, theoretically, require higher spatial sampling rates. Hence, it is important to verify whether the non-directional beam patterns observed in Figs. \ref{f-beam-pat-sim}b and \ref{f-beam-pat-sim}d are results of spatial undersampling.

\begin{lemma}\label{lmm-1}
For a uniform rectangular sampling grid with spacing $d_1$ in the $x$-direction and $d_2$ in the $y$-direction, an EM signal can be recovered from the data sampled on this grid if its spatial frequencies in the $x$- and $y$-directions satisfy $|k_x|<k_{s,x}$ and $|k_y|<k_{s,y}$, where $k_{s,x}=\frac{\pi}{d_1}$ and $k_{s,y}=\frac{\pi}{d_2}$ are the maximum supported frequencies in the two directions.
\end{lemma}

\begin{figure}[b]
\vspace{-1.5em}
\centering
\subfloat[]{
\includegraphics[width=0.19 \textwidth, trim=0.3cm 0cm 0cm 0.05cm, clip]{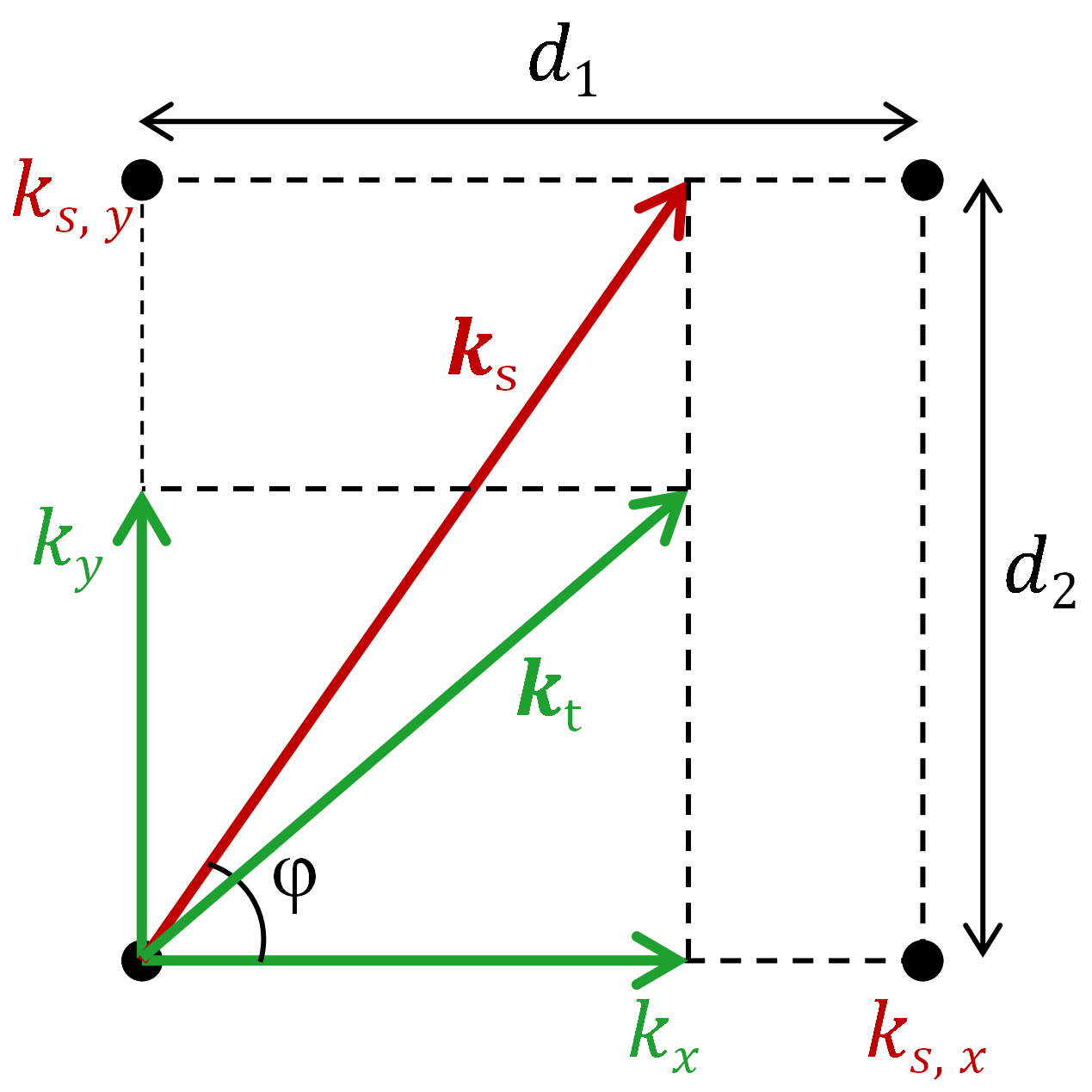}}
\subfloat[]{
\includegraphics[width=0.26 \textwidth]{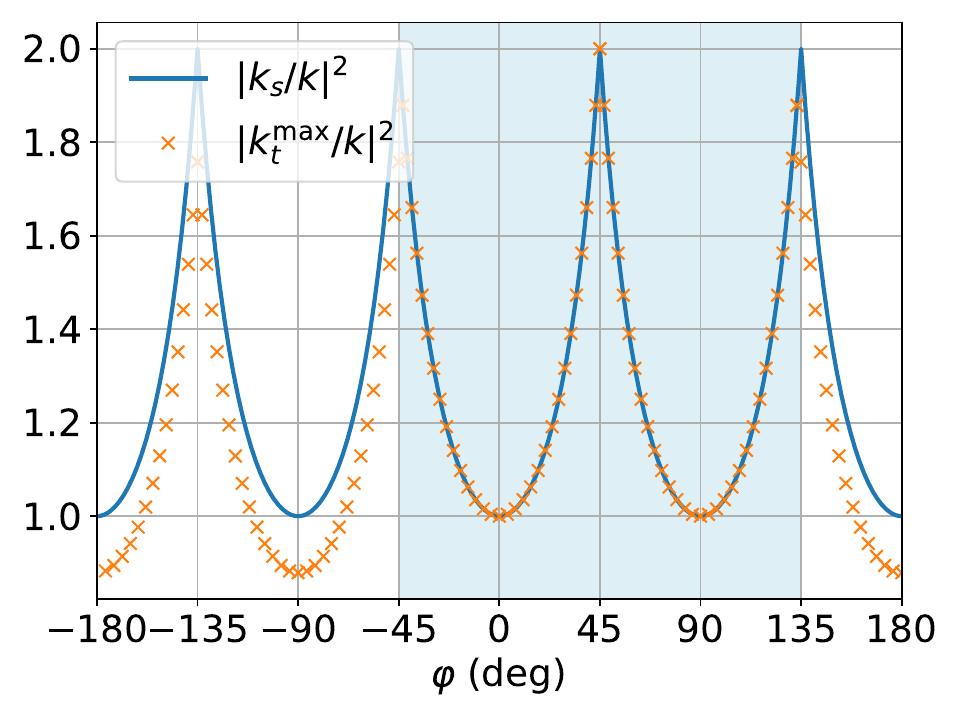}}
\caption{(a) Illustration of the spatial frequencies supported by a rectangular grid. (b) Maximum spatial frequency supported by a UPA with $\alpha_1=\alpha_2=0.5$ (solid line), along with the spatial frequencies of the outermost codewords shown in Fig. \ref{f-hf-area}b (orange ``x'').}
\label{f-max-k}
\end{figure}

\begin{IEEEproof}
The Nyquist criterion states that a uniform sampling with spacing $d$ can support signals with wavelengths $\lambda \geq 2d$, where \textit{support} means that the signal can be accurately recovered from the sampled data. Accordingly, as shown in Fig. \ref{f-max-k}a, the spatial frequencies supported by a uniform rectangular sampling grid in the $x$- and $y$-directions are constrained by the spacings $d_1$ and $d_2$, respectively. Specifically, the apparent wavelength in the $x$-direction, $\lambda_x=\frac{2\pi}{|k_x|}$, must fulfill $\lambda_x\geq 2d_1$, which gives $|k_x| \leq \frac{\pi}{d_1}$. Similarly, we have $|k_y| \leq \frac{\pi}{d_2}$. Since $k_x$ and $k_y$ can be negative values in spatial signals, to ensure alias-free recovery, the equality in $|k_x| \leq \frac{\pi}{d_1}$ and $|k_y| \leq \frac{\pi}{d_2}$ should be excluded, leading to the strict conditions $|k_x| < k_{s,x}$ and $|k_y| < k_{s,y}$.
\end{IEEEproof}

\begin{figure}[t]
\centering
\subfloat[]{
\includegraphics[width=0.23 \textwidth, trim=2.6cm 0.5cm 0.6cm 2.9cm, clip]{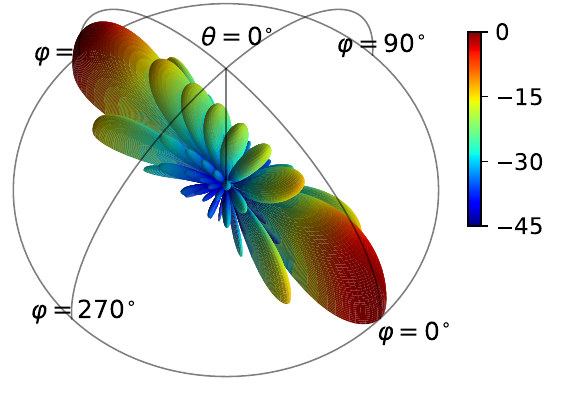}}\;
\subfloat[]{
\includegraphics[width=0.23 \textwidth, trim=2.6cm 0.5cm 0.6cm 2.9cm, clip]{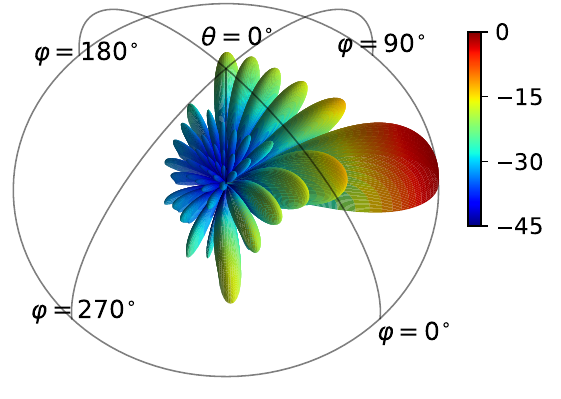}}\\
\subfloat[]{
\includegraphics[width=0.23 \textwidth, trim=2.6cm 0.8cm 0.6cm 3cm, clip]{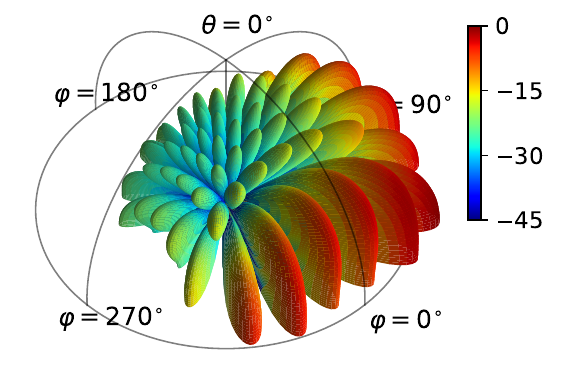}}\;
\subfloat[]{
\includegraphics[width=0.23 \textwidth, trim=2.6cm 0.8cm 0.6cm 3cm, clip]{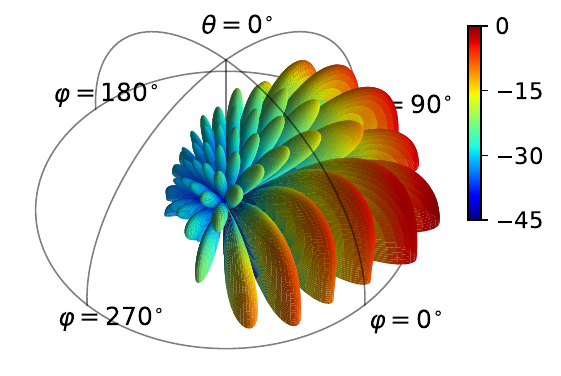}}
\caption{Beam patterns of precodings (a) $\mathbf{w}_1$ and (b) $\mathbf{w}_2$; Beam patterns of precodings with the same phase gradient as the evanescent codeword $\boldsymbol{v}_{14,16}$ on UPAs with (c) $\alpha_1=\alpha_2=0.4$ and (d) $\alpha_1=\alpha_2=0.25$. Isotropic radiation pattern is assumed.}
\label{f-beam-pat-sim2}
\vspace{-0.5em}
\end{figure}

Lemma \ref{lmm-1} can also be explained by the relation between the sampling matrix $\mathbf{Q}$ and the periodicity matrix $\mathbf{P}$ depicted by Eq. (10) in \cite{pizzo22}. As illustrated in Fig. \ref{f-max-k}a, the spatial frequency supported at different propagating direction is determined by the length of the vector $\boldsymbol{k}_s$. Obviously, the diagonal directions accommodate the highest spatial frequency while the two orthogonal directions along the grid support the lowest spatial frequencies, as shown in Fig. \ref{f-max-k}b. Since the grid has a larger sampling spacing in diagonal directions, this conclusion is quite counter-intuitive at first glance, but it can be easily verified by beamforming simulations. As detailed in Table \ref{t-ab}, two precoding vectors, representing EM waves with the same spatial frequency $|\boldsymbol{k}_t|=k$ but propagating towards $\varphi=0^{\circ}$ and $\varphi=45^{\circ}$, respectively, are selected for the verification. The simulations are based on a UPA with 8$\times$8 half-wavelength antennas. As shown in Figs. \ref{f-beam-pat-sim2}a and \ref{f-beam-pat-sim2}b, $\mathbf{w}_1$ generates a beam with a grating lobe in the opposite direction while only the main lobe appears for $\mathbf{w}_2$, which proves that a higher spatial frequency is supported in the diagonal direction of a UPA.
 
\begin{table}[!h]
\centering
\caption{Parameters of precoding vectors $\mathbf{w}_1$ and $\mathbf{w}_2$}
\label{t-ab}
\renewcommand\arraystretch{1.2}
\begin{tabular}{|c|c|c|c|c|}
\hline
Precoding & $k_x$ & $k_y$ & $|\boldsymbol{k}_t|$ & Beam direction \\ \hline
$\mathbf{w}_1$    & $k$  & 0  & $k$  & $\theta=90^{\circ},\varphi=0^{\circ}$      \\ \hline
$\mathbf{w}_2$    & $\frac{\sqrt{2}}{2}k$ & $\frac{\sqrt{2}}{2}k$ & $k$  & $\theta=90^{\circ},\varphi=45^{\circ}$    \\ \hline
\end{tabular}
\end{table}

According to \eqref{eq-kxky}, the codewords in the outermost grid of Fig. \ref{f-hf-area}b have higher spatial frequencies than the inner ones in their respective azimuth directions with the codeword at the upper-right corner has the highest spatial frequency. If a UPA supports the spatial frequencies of these outermost codewords, it can also supports the spatial frequencies of all other codewords.

\begin{corollary}\label{coro-3}
A UPA equipped with $N_1\times N_2$ half-wavelength antennas supports the spatial frequency of a codeword $\boldsymbol{v}_{lm}$ in $\mathbf{CB}_{N_1,N_2}^{O_1,O_2}$ if $l\neq\frac{N_1O_1}{2}$ or $m\neq\frac{N_2O_2}{2}$.
\end{corollary}
\begin{IEEEproof}
According to Lemma \ref{lmm-1}, for a UPA with $\alpha_1=\alpha_2=0.5$, the supported spatial frequency $\boldsymbol{k}_t=(k_x,k_y)$ must satisfy $|k_x|< \frac{\pi}{0.5\lambda}=k$ and $|k_y| < \frac{\pi}{0.5\lambda}=k$. Acoording to \eqref{eq-kxky}, these two conditions are violated only when $l\neq\frac{N_1O_1}{2}$ or $m\neq\frac{N_2O_2}{2}$. Consequently, a UPA with half-wavelength antenna spacing does not support the spatial frequencies of codewords with indices $l=\frac{N_1O_1}{2}$ or $m = \frac{N_2O_2}{2}$. The spatial frequencies of all other codewords are supported.
\end{IEEEproof}

In Fig. \ref{f-hf-area}b, the codewords located in the rightmost column and uppermost row correspond to codewords with indices $l=\frac{N_1O_1}{2}$ or $m=\frac{N_2O_2}{2}$, representing wave vectors in the azimuth range $(-45^{\circ},135^{\circ})$. As shown in  Fig. \ref{f-max-k}b, the spatial frequencies of these codewords are equal to $|\boldsymbol{k}_s|$ in their respective azimuths, indicating that the spatial frequencies of these codewords are not supported. The codeword $\boldsymbol{v}_{14,16}$ in $\mathbf{CB}_{8,8}^{4,4}$ satisfies $m=\frac{N_2O_2}{2}$, thus the simulation results in \ref{f-beam-pat-sim}b and \ref{f-beam-pat-sim}d should include grating lobes due to spatial undersampling.

To verify this, a precoding vector with the same spatial frequency as $\boldsymbol{v}_{14,16}$ was applied to a UPA with smaller antenna spacing for the beam pattern simulation. As shown in Fig. \ref{f-beam-pat-sim2}c, when the antenna spacing is reduced to $0.4\lambda$, the beam lobes in the azimuth range $(135^{\circ},315^{\circ})$ vanish compared to Fig. \ref{f-beam-pat-sim}b, indicating the elimination of grating lobes. Further diminishing the antenna spacing to $0.25\lambda$ only slightly lowers the side lobe level and does not improve total radiation efficiency, as shown in Fig. \ref{f-beam-pat-sim2}d.

It is now evident that the diffusive beam pattern and low radiation efficiency of an evanescent codeword is not caused by spatial undersampling. Indeed, the evanescent codewords are incapable of generating directional beams in physical space, even with a denser array. Note that, if a codeword is directly loaded on a UPA, its actual spatial frequency varies with the antenna spacing. As a result, the spatial frequencies of the codewords with indices $l=\frac{N_1O_1}{2}$ or $m=\frac{N_2O_2}{2}$ always satisfy $|k_x|=k_{s,x}$ and $|k_y|=k_{s,y}$ on any UPAs. In other words, the spatial frequencies of these codewords are not supported by any UPAs, regardless of the antenna spacing.

\subsection{Codebook-based Representation of Multi-path Channels}
We have demonstrated that evanescent codewords are redundant in array beamforming. However, the actual wireless channel in a rich scattering environment is more complex than the LOS channel described by \eqref{eq-h}. It is important to investigate whether evanescent codewords are required for multi-path channel representation in CSI reporting. In TR 38.901 \cite{38901r16}, the non-line-of-sight (NLOS) channel is modeled as a summation of clusters, each comprising multiple rays within specific angular spreads.

Importantly, the channel generation procedure specified in TR 38.901 explicitly demands that each ray is associated with real values of $\theta$ and $\varphi$, ensuring that no evanescent components are present. Consequently, the phase response of each ray is equivalent to the far-field LOS model \eqref{eq-h}. This NLOS model aligns with Theorem 2 in \cite{pizzo22b}, which demonstrates that a wireless channel can be approximated by a Fourier plane-wave series expansion. Note that, ``plane-wave'' indicates that the expansion includes only propagating waves with spatial frequencies no higher than the free-space one, $k$. Consequently, the multi-path far-field channel acts as a low-pass filter, inherently excluding evanescent waves.

\begin{figure}[t]\label{fig-s2}
\centering
\includegraphics[width=0.48 \textwidth]{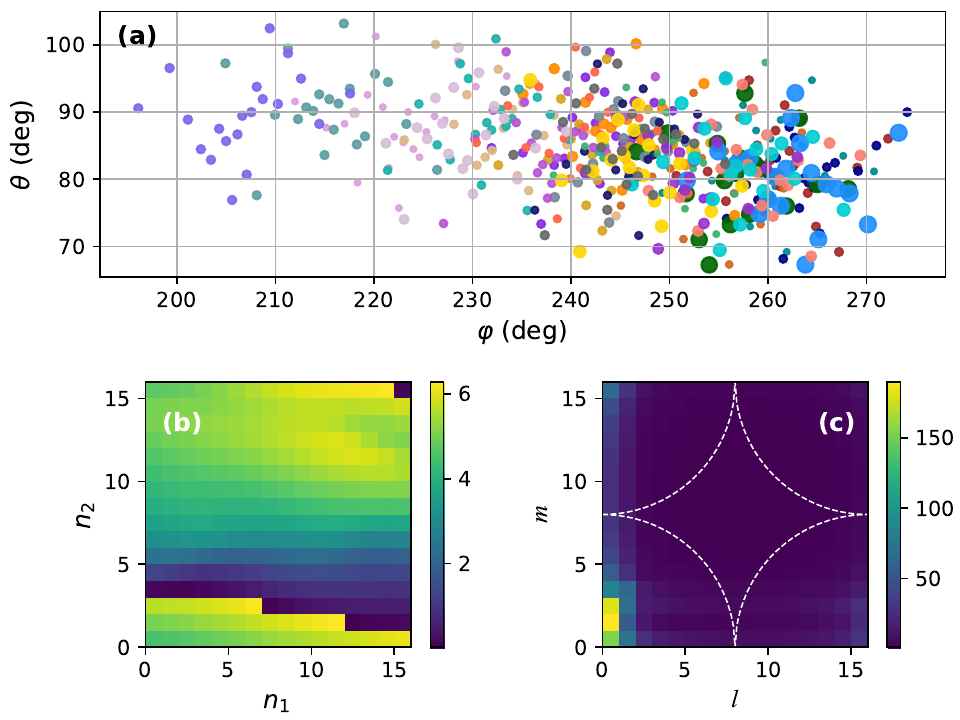}
\caption{(a) Angles of departure of rays in a 3GPP far-field channel; (b) phase distribution of the channel; (c) projection of the channel onto the codebook $\mathbf{CB}_{16,16}^{1,1}$. The patch size in (a) indicates the relative amplitude response. The white dashed curve in (c) depicts the zone boundary $\Gamma$.}
\vspace{-0.8em}
\end{figure}

To verify this, a far-field channel between a base station (BS) with a 16$\times$16 half-wavelength antenna array and a single-antenna UE in a scattering-rich indoor environment was generated, following TR 38.901. The channel consisted of 20 clusters, each containing 25 rays. Fig. \ref{fig-s2}a shows the angles of departure of each ray, with patch size indicating the amplitude response. In Fig. \ref{fig-s2}b, some randomness in the phase distribution, resulting from the summation of multiple paths, can be observed. However, projecting the channel onto the codebook $\mathbf{CB}_{16,16}^{1,1}$ reveals that the channel energy is mainly concentrated in the propagating zone, as shown in Fig. \ref{fig-s2}c. Hence, the representation of multi-path channel does not requires evanescent codewords.

On the other hand, if the summation of multiple paths could lead to a composite channel containing both propagating and evanescent components, it would suggest that the linear combination of certain plane waves forms an evanescent wave. This would contradict the sampling theorem, implying that high-frequency components can be reconstructed from data sampled at a rate below the Nyquist limit. In summary, due to the low-pass nature of far-field channels, evanescent codewords are unnecessary for channel representation, rendering them redundant for CSI feedback.
\vspace{-0.5em}
\subsection{The Evanescent Zone in Wideband Communications}\label{sec-wb}
According to Corollary \ref{coro-1}, an evanescent codeword may transform into a propagating one when applied to a UPA with larger antenna spacing at the same frequency. In wideband transmission, this transformation can also occur on a UPA with fixed antenna spacing, as the zone boundary $\Gamma$ is determined by the normalized antenna spacing. Specifically, the normalized antenna spacing varies across subcarriers because it is frequency-dependent: $\alpha_1(f)=\frac{d_1}{\lambda(f)}$ and $\alpha_2(f)=\frac{d_2}{\lambda(f)}$, where $f$ denotes the central frequency of the subcarrier.

\begin{figure}[h]
\centering
\includegraphics[width=0.24\textwidth]{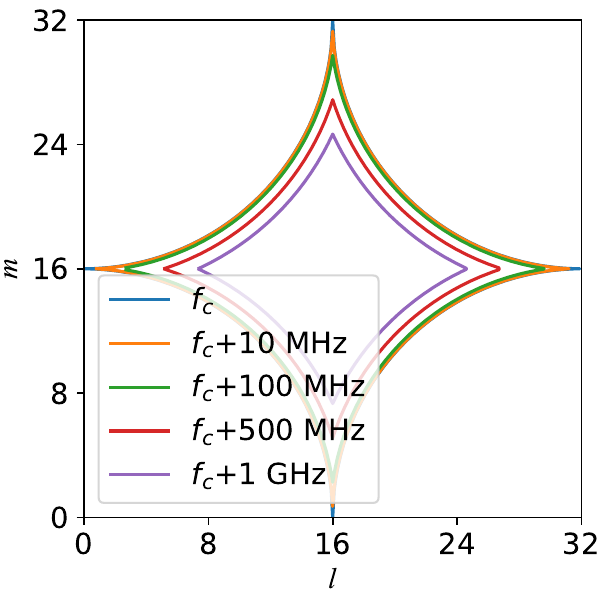}
\caption{The zone boundary $\Gamma$ in $\mathbf{CB}_{8,8}^{4,4}$ for different sub-carriers. The antenna spacing is $0.5\lambda$ at central frequency $f_c$ = 10 GHz.}
\label{f-evanescent-zones}
\vspace{-1em}
\end{figure}

As shown in Fig. \ref{f-evanescent-zones}, for a codebook designed for a fixed UPA, the evanescent zone contracts as the offset between the subcarrier frequency and the central frequency increases. Table \ref{t-redundancy} provides the number of evanescent codewords in $\mathbf{CB}_{8,8}^{4,4}$ at various frequency offsets from the central frequency. While the evanescent zone remains nearly unchanged within a 75 MHz bandwidth, a wider band requires careful consideration to account for the evolving evanescent zone when addressing redundancy reduction.

\begin{table}[h]
\centering
\caption{Numbers of evanescent codewords and redundancy for different sub-carriers at 10 GHz band}
\renewcommand\arraystretch{1.2}
\begin{tabular}{|c|c|>{\centering\arraybackslash}p{1cm}|c|}
\hline
$f-f_c$& $\alpha_1,\alpha_2$ & $N_{\text{eva}}$ & Redundancy $r_{\text{eva}}$ \\
\hline
0 MHz & 0.50000 & 229 & 22.36\% \\
\hline
15 MHz & 0.50075 & 229 & 22.36\% \\
\hline
30 MHz & 0.50150 & 225 & 21.97\% \\
\hline
75 MHz & 0.50375 & 225 & 21.97\% \\
\hline
100 MHz & 0.50500 & 205 & 20.02\% \\
\hline
400 MHz & 0.52000 & 173 & 16.89\% \\
\hline
1 GHz & 0.55000 & 117 & 11.43\% \\
\hline
\end{tabular}
\label{t-redundancy}
\vspace{0.3em}
\end{table}

\section{Practical Codebook Optimization}\label{sec-4}
\subsection{Codebook Compression}
In previous sections, we thoroughly discussed the presence of redundant codewords in Kronecker product-based codebooks. If this redundancy is not accounted for in the standard design, it can result in unnecessary signaling overhead and increased operational complexity in MIMO communications. To address this issue, we propose a method to compress the codebooks based on Proposition \ref{prop-1}. The compression involves excluding the evanescent codewords and reordering the propagating codewords, as detailed in Algorithm \ref{alg-1}.

\begin{algorithm}
\caption{Compression of a Kronecker product-based codebook}
\begin{algorithmic}[1]\label{alg-1}
\REQUIRE $N_1$, $N_2$, $O_1$, $O_2$, $\alpha_1$, $\alpha_2$
\ENSURE The list of indices of propagating codewords $R$
\STATE Initialize $R \gets []$
\FOR{$l \gets 0$ to $N_1O_1-1$}
	\STATE $t_1=l\leq\frac{N_1O_1}{2} \;?\; \frac{l}{\alpha_1N_1O_1} : \frac{l-N_1O_1}{\alpha_1N_1O_1}$
    \FOR{$m \gets 0$ to $N_2O_2-1$}
    	\STATE $t_2=m\leq\frac{N_2O_2}{2}\;?\; \frac{m}{\alpha_2N_2O_2}:\frac{m-N_2O_2}{\alpha_2N_2O_2}$
        \IF{$t_1^2+t_2^2\leq 1$}
            \STATE Append ($l$,$m$) to $R$
        \ENDIF
    \ENDFOR
\ENDFOR
\RETURN $R$
\end{algorithmic}
\end{algorithm}

In the current NR standard, if a codeword $\boldsymbol{v}_{lm}$ is selected to represent the channel state, its index $(l,m)$ is reported. However, if codebook compression is applied according to Algorithm \ref{alg-1}, the index corresponding to $(l,m)$ in the list $R$ is reported instead. The compression does not alter the structure of Type I, Type II or eType II codewords, nor does it modify the SD bases. Consequently, the compression does not impact the PMI-based feedback procedure. The only additional operation required is a simple mapping between the reported PMI and the index $(l,m)$, which can be efficiently performed using a lookup in the list $R$.

\subsection{Overhead and Efficiency Analysis}
Compressing codebooks by excluding evanescent codewords proves advantageous for codebook-based CSI feedback in MIMO communications. Firstly, the CBSR signaling overhead can be significantly reduced without requiring additional processing. As specified in TR 38.214 \cite{38214r16}, the CBSR signaling for Type I feedback consists of a bit sequence of length $N_{\text{tot}}=N_1O_1N_2O_2$. For Type II or eType II feedback, the sequence length becomes $\lceil\log_2 \binom{O_1O_2}{4}\rceil+8N_1N_2$. After compressing the codebook, the bit sequence length is reduced to $(N_{\text{tot}}-N_{\text{eva}})$ for Type I feedback and $\lceil\log_2 \binom{O_1O_2}{4}\rceil+8\frac{N_{\text{tot}}-N_{\text{eva}}}{O_1O_2}$ for (e)Type II feedback, respectively.

Secondly, the compression can reduce the overhead in PMI reporting. According to TR 38.212 \cite{38212r16}, this compression impacts the overhead of the indicators $i_{1,1}$ and $i_{1,2}$ for Type I single-panel (SP) feedback (Table 6.3.1.1.2-1) and $i_{1,2}$ for Type II feedback (Table 6.3.2.1.2-1). The overhead reduction in CBSR signaling and PMI reporting for different arrays is summarized in Table \ref{t-overhead}. For both Type I SP and Type II feedback, CBSR signaling overhead is reduced by over 22\% across various array sizes. Since CBSR is a UE-specific configuration, reducing signaling overhead conserves more channel resources for data transmission. However, the compression has minimal effect on CSI feedback overhead. The observed 2-bit reduction in Type II feedback arises from reporting $i_{1,2}$ via combination numbers.

\begin{table}[b]
\vspace{-0.5em}
\centering
\caption{Overhead reductions achieved through codebook compression}
\renewcommand\arraystretch{1.2}
\setlength{\tabcolsep}{3pt}
\begin{tabular}{|c|c|cc|cc|}
\hline
\multirow{2}{*}{$N_1\times N_2$} & \multirow{2}{*}{$N_{\text{tot}}$ / $N_{\text{eva}}$} & \multicolumn{2}{c|}{\tabincell{c}{CBSR overhead reduction\\(bits / percentage)}}       & \multicolumn{2}{c|}{\tabincell{c}{Feedback overhead\\reduction (bits)}} \\ \cline{3-6} 
&                    & \multicolumn{1}{c|}{Type I SP} & Type II & \multicolumn{1}{c|}{Type I SP} & \multicolumn{1}{c|}{Type II} \\ \hline
4$\times$4 & 256 / 61 & \multicolumn{1}{c|}{61 / 24\%}  & 40 / 30\% & \multicolumn{1}{c|}{0}  & 2                       \\ \hline
8$\times$8 & 1024 / 229 & \multicolumn{1}{c|}{229 / 22\%}  & 136 / 26\%  & \multicolumn{1}{c|}{0}  & 2                       \\ \hline
16$\times$16 & 4096 / 889 & \multicolumn{1}{c|}{889 / 22\%}  & 488 / 24\%  & \multicolumn{1}{c|}{0}  & 2 \\ \hline
32$\times$32 & 16384 / 3533 & \multicolumn{1}{c|}{3533 / 22\%}  & 1832 / 22\%  & \multicolumn{1}{c|}{0}  & 2                       \\ \hline
\end{tabular}\label{t-overhead}
\vspace{0.5em}
\par \small \raggedright Note: $O_1=O_2=4$ and $\alpha_1=\alpha_2=0.5$ are applied. Four SD bases are selected in Type II feedback.
\end{table}

Codebook-based beam training has been applied in millimeter-wave MIMO communications involving large arrays. When exhaustive search is employed for the training, the associated pilot overhead becomes prohibitively high. For example, according to Table \ref{t-overhead}, the beam training for a UPA with 32$\times$32 antennas requires transmitting 16384 beams. Although new algorithms, such as hash-based multi-arm beam training \cite{wangx22, xuy24}, have been proposed to accelerate the training process, pilot overhead remains a critical challenge for ELAA. Compressing the codebook offers a practical solution to significantly reduce the pilot overhead. When integrated with hash-based beam training, the pilot overhead reduction becomes even more pronounced. According to \cite{xuy24}, the number of pilots required to achieve a training accuracy of $1-1/M_s$ is given by $\lceil \frac{N_{\text{prop}}}{B}\rceil\log M_s$, where $B$ is the number of arms. Assuming $B=16$ and $M_s=100$ (corresponding to 99\% identification accuracy), the codebook compression reduces the pilot overhead from 5120 to 4020, achieving a 21.5\% reduction.

Beyond overhead reduction, the proposed codebook compression also decreases channel estimation complexity. For codebook-based CSI feedback, the computation demand for SD bases selection can be significantly reduced due to the compression. In hash-based beam training, fewer reference signals need to be received and processed at the UE, boosting the training process. Based on the above analysis, the pilot reduction directly improves training efficiency by 21.5\%. Accordingly, system performance can be enhanced with the compression of the codebooks.

As discussed in Section \ref{sec-wb}, the evanescent zone within a codebook varies according to carrier frequencies. Therefore, the compression of a codebook should be executed based on the allocated transmission bandwidth. If the compression is based on the central frequency, some propagating codewords for other subcarriers may be unintentionally excluded, potentially leading to performance degradation in wideband transmission. The most straightforward solution involves identifying the maximum propagating codeword set $R_{\max}$ by Algorithm \ref{alg-1} for a specified bandwidth. A subband-based codebook compression can achieve a more accurate exclusion of evanescent codewords but may also introduce additional complexity and signaling overhead. In 5G system, the band width part (BWP) activated for a UE is usually much narrower (typically 5 to 10 MHz) than the whole bandwidth supported. Hence, the subbands usually share a same evanescent zone and the exclusion of evanescent codewords does not need to be done as per subband.

\subsection{System-Level Simulations}\label{sec-sys-sim}

To verify the feasibility of codebook compression in MIMO communications, system-level simulations were conducted in our NR platform. The 3GPP indoor-office scenario was chosen for the simulation as it offers more complicated wireless scattering conditions, particularly involving paths uniformly distributed across the whole azimuth range. As depicted in Fig.7.2-1 and Table 7.2-2 in TR 38.901 \cite{38901r16}, 12 BSs (i.e., gNB), are mounted on the office ceiling, 3 m above the ground and 20 m apart. The office (120 m$\times$50 m) is partitioned into 12 sectors and each gNB serves one sector that contains 15 uniformly distributed UEs. The UEs, equipped with 2$\times$2 dual-polarized half-wavelength antennas, are moving at a height of 1 m from the ground. The detailed parameters for the simulations are summarized in Table \ref{t-sim-setup}. The following setup were also applied in the simulations: (1) eType II codebook-based CSI feedback was utilized; (2) PMIs were reported on resource block (RB) granularity; (3) FTP traffic model 1 \cite{36814r9} was chosen for MU transmission; (4) the modulation coding scheme (MCS) and the number of transmission layers were determined adaptively based on the channel condition.

\begin{table}[h!]
\centering
\caption{Setup for system-level simulations}
\label{t-sim-setup}
\renewcommand\arraystretch{1.2}
\begin{tabular}{|p{0.16\textwidth}|p{0.28\textwidth}|}
\hline
\multicolumn{1}{|c|}{\textbf{Parameters}}& \multicolumn{1}{c|}{\textbf{Values}}                                                 \\ \hline
Carrier Frequency                     & 6 GHz                                                                       \\ \hline
Channel Model                         & 3GPP TR 38.901 indoor-office setup \cite{38901r16}                                    \\ \hline
Antenna setup and port layouts at gNB &(M, N, P, Mg, Ng, Mp, Np) = (8,8,2,1,1,8,8), (4,8,2,1,1,4,8), (4,4,2,1,1,4,4)
\newline (dH, dV) = (0.5 $\lambda$, 0.5 $\lambda$) \\ \hline
Antenna setup and port layouts at UE  & (M, N, P, Mg, Ng, Mp, Np) =(1,2,2,1,1,1,2)\newline (dH, dV) = (0.5 $\lambda$, 0.5 $\lambda$)            \\ \hline
Modulation                            & Up to 256 QAM                                                               \\ \hline
Tx power                          & 44 dBm (gNB), 23 dBm (UE)                                                                     \\ \hline
Noise figure      & 5 dB (gNB), 9 dB (UE)      
\\ \hline
Numerology                            & 14 OFDM symbol per slot, 15 kHz SCS                                         \\ \hline
UE Bandwidth                          & 10 MHz, 52 RBs                                                               \\ \hline
UE reception                          & MMSE-IRC \cite{tavares14}                                                              \\ \hline
UE moving speed                       & $\leq$ 3 km/h                                                                   \\ \hline
Network Layout                        & 12 sectors, 15 UEs per sector                                     \\ \hline
CSI feedback delay                    & 5 ms                                                                        \\ \hline
MIMO scheme                           & MU-MIMO with rank = 1-4 per UE                                            \\ \hline
\end{tabular}
\end{table}

\begin{figure}[h]
\centering
\subfloat[]{
\includegraphics[width=0.2 \textwidth, trim=0.16cm 0.05cm 0cm 0.0cm, clip]{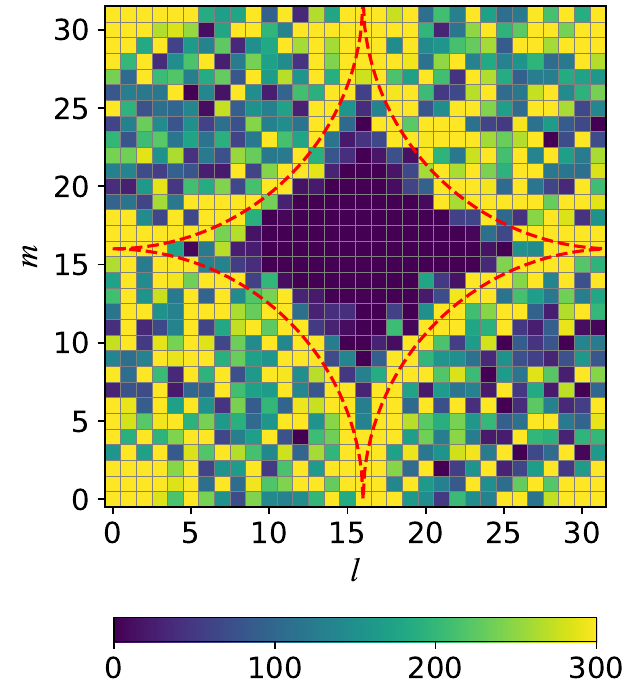}}\;
\subfloat[]{
\includegraphics[width=0.26 \textwidth]{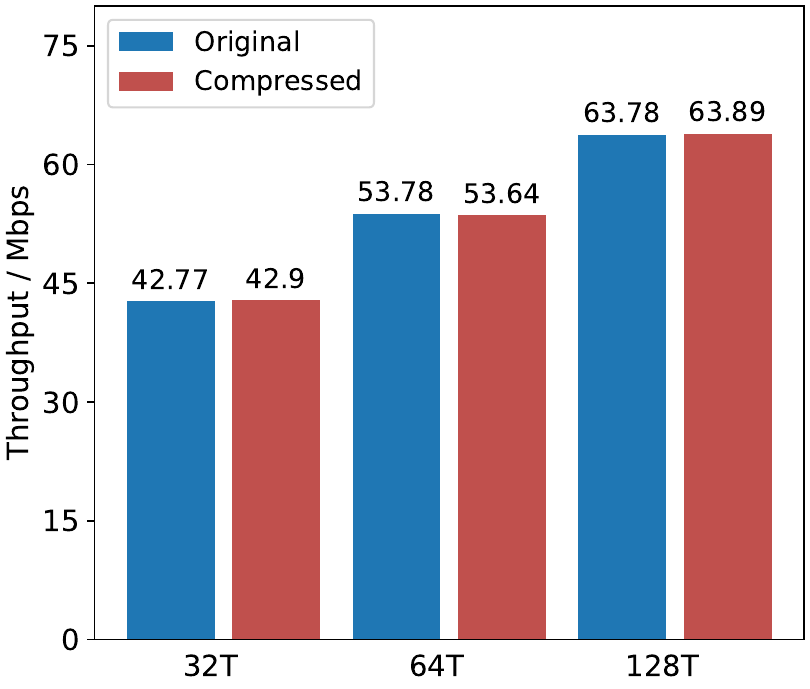}}\\
\subfloat[]{
\includegraphics[width=0.2 \textwidth, trim=0.16cm 0.05cm 0cm 0.0cm, clip]{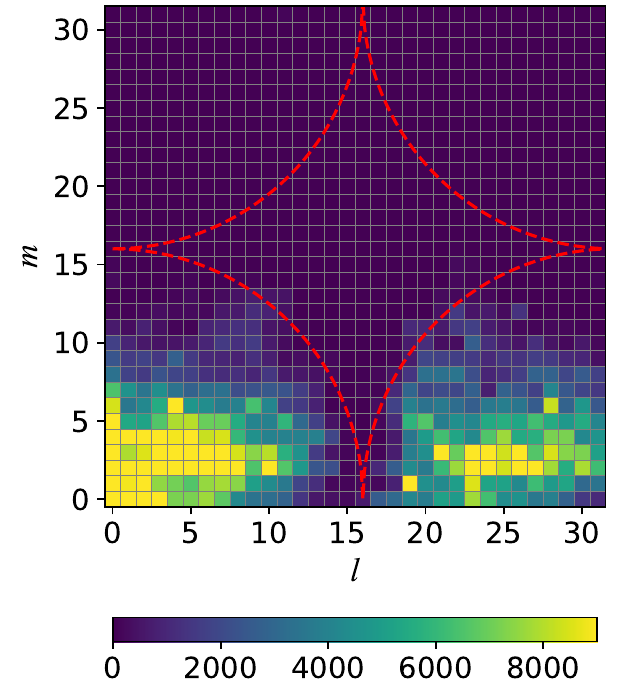}}\;
\subfloat[]{
\includegraphics[width=0.26 \textwidth]{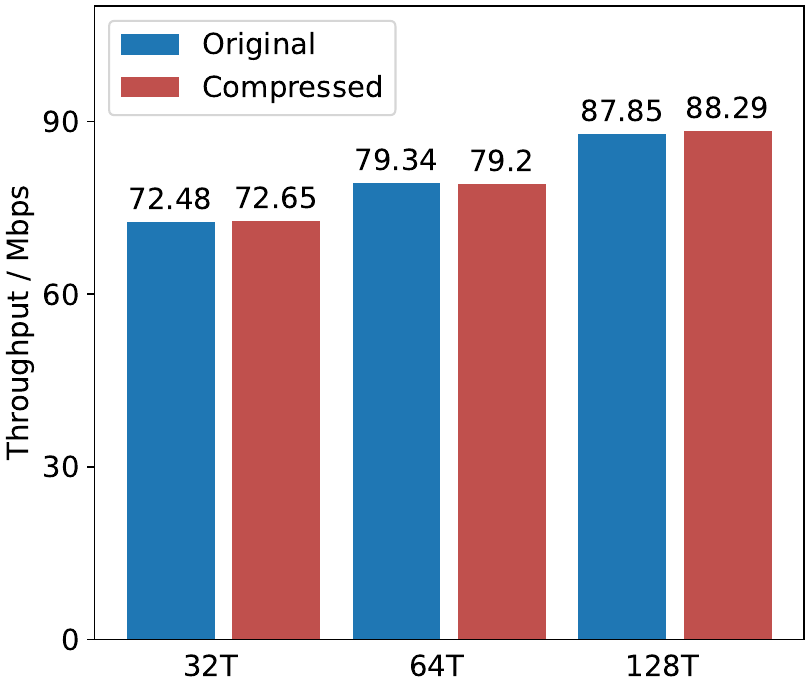}}
\caption{(a) Codeword selection counts and (b) system throughput in the indoor-office scenario; (c) codeword selection counts and (d) system throughput in the UMi scenario. The dashed curves in (a) and (c) represent the zone boundary.}
\label{f-sim-result}
\vspace{-0.5em}
\end{figure}

First, the uncompressed codebook $\mathbf{CB}_{8,8}^{4,4}$ was employed for BSs equipped with a dual-polarized 8$\times$8 array (128T). As shown in Fig. \ref{f-sim-result}a, the codewords selected during the simulation were predominantly located within the propagating zone. To improve visualization, selection counts exceeding 300 were capped at 300 in the heatmap, allowing better contrast for codewords that were never selected. Theoretically, no evanescent codewords should be selected; however, we observed that some evanescent codewords were selected near the zone boundary $\Gamma$. There were a total of 253,280 selections with approximately 13.5\% being the evanescent codewords.

Next, we compared the system throughput with and without codebook compression under three dual-polarized BS array configurations: 1) $N_1=N_2=4$ (32T); 2) $N_1=4$, $N_2=8$ (64T); and (3) $N_1=N_2=8$ (128T). As shown in Fig. \ref{f-sim-result}b, the throughput remained nearly identical regardless of whether evanescent codewords were included in the codebook.

We have fully addressed the redundancy of evanescent codewords, yet some of them were selected in the simulations. The following factors should account for the selections.

\begin{figure}
\centering
\subfloat[]{
\includegraphics[width=0.23 \textwidth]{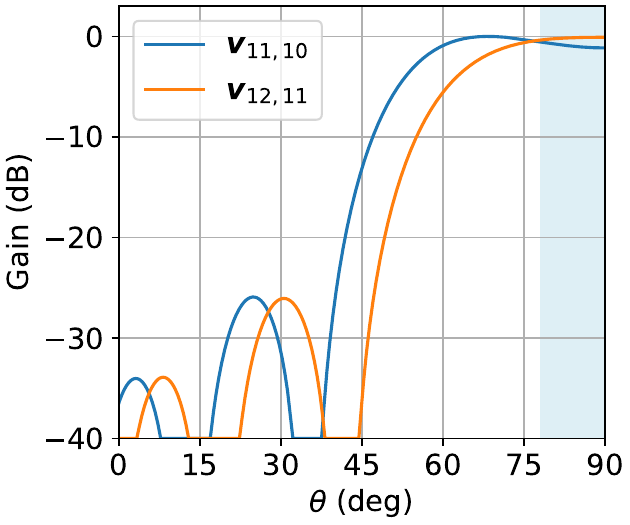}}\quad\;
\subfloat[]{
\includegraphics[width=0.2 \textwidth]{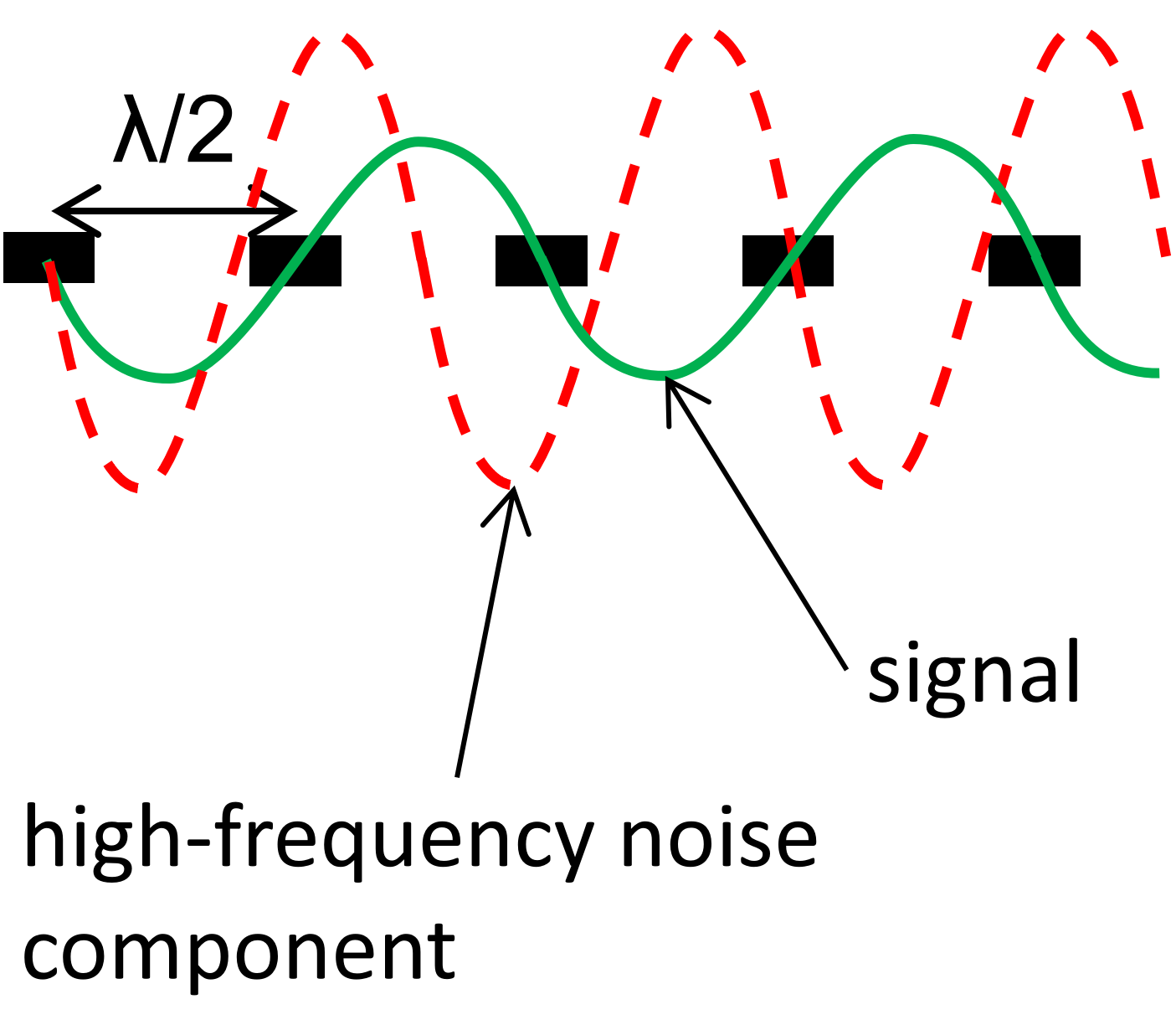}}\\
\subfloat[]{
\includegraphics[width=0.48 \textwidth]{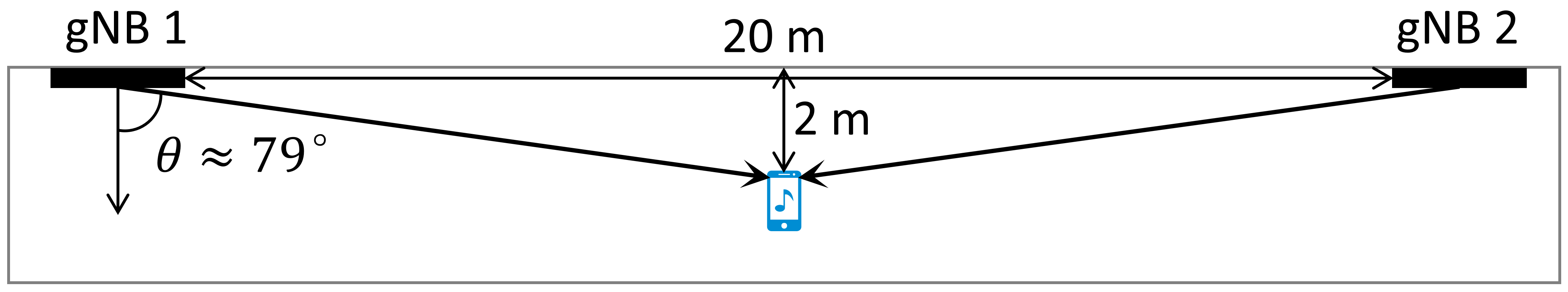}}
\caption{(a) beam patterns of $\boldsymbol{v}_{12,11}$ and $\boldsymbol{v}_{11,10}$; (b) the signal and a high frequency noise component on an array; (c) Schematic diagram of a downlink transmission near the cell edge.}
\label{f-ue-pos}
\vspace{-0.5em}
\end{figure}

\begin{enumerate}
\item \textbf{Highly Oblique Paths}:
When a highly oblique path (i.e., $\theta \to \frac{\pi}{2}$) exists between a gNB and a UE, a codeword near the zone boundary is more likely to match the channel. Although the beam peak of an evanescent codeword is not observable, its side lobes or even part of its main lobe can project into the physical space due to the finite array aperture. This effect is particularly pronounced for evanescent codewords near the zone boundary. In Fig. \ref{f-ue-pos}a, the main lobe generated by the evanescent codeword $\boldsymbol{v}_{12,11}$ from $\mathbf{CB}_{8,8}^{4,4}$ partially extends into the physical space. When a UE is located in the light blue region (i.e., $\theta > 78^{\circ}$), $\boldsymbol{v}_{12,11}$ will be selected instead of the adjacent propagating codeword $\boldsymbol{v}_{11,10}$. However, the beamforming gains of both codewords are so close in this region that selecting either one has a negligible impact on throughput and the choice of codeword can be easily influenced by noise.
\item \textbf{Interference}: 
In the presence of interference from other users or neighboring cells, the NR system tends to select a codeword adjacent to the one indicated by channel measurements to suppress interference. If the intended codeword is near the zone boundary, strong interference may result in the selection of an adjacent evanescent codeword.
\item \textbf{Broadband Noise}:
Transmission signals in MIMO systems are usually accompanied by broadband noise. In low-SNR situations, high-frequency noise components may have larger amplitudes than the signal, as depicted in Fig. \ref{f-ue-pos}b. The codeword selection algorithm may then favor an evanescent codeword that aligns with the high-frequency noise component. Consequently, the system is more prone to mistakenly select an evanescent codeword in low-SNR environments. In this sense, the exclusion of evanescent codewords can help suppress high-frequency noises.
\end{enumerate}

In certain rounds of the indoor simulations, we examined the locations of UEs whose CSI report included evanescent codewords. We found that evanescent codewords were most likely to be selected when UEs were located near cell edges, as depicted in Fig. \ref{f-ue-pos}c. This observation is reasonable for several reasons. First, the path between the gNB and UE is nearly horizontal near cell edges, i.e., highly oblique. Second, inter-cell interference is the strongest in this region, leading to extremely low SINR. Moreover, the radiation efficiency of an array is extremely low in highly oblique direction \cite{38901r16}, further degrading the SINR. Consequently, the impacts of the above three factors culminate at cell edges, increasing the likelihood of selecting evanescent codewords near the zone boundary.

Clearly, the selection of evanescent codewords primarily occurred in low-SINR regions. Although excluding evanescent codewords may have improved the accuracy of the reported CSI and suppressed high-frequency noise components, it did not alter the underlying channel conditions. As a result, no direct throughput enhancement was achieved by codebook compression. Nevertheless, evanescent codewords are redundant, as their exclusion did not degrade throughput.

We further extended the simulations to the UMi scenario \cite[Table 7.2-1]{38901r16} to examine the selection of evanescent codewords. As shown in Fig. \ref{f-sim-result}c, unlike the indoor-office simulations, the selected codewords were primarily located in the lower part of the codebook. There were a total of 1,388,762 codeword selections and only 8550 evanescent codewords were selected. In this scenario, the UEs were distributed within the angle range $\theta\in[-60^{\circ}, 60^{\circ}]$, hence there were less highly oblique paths between the BS and UEs, leading to a much lower probability of evanescent codeword selection. Similarly, compressing the codebook had no noticeable impact on throughput, as illustrated in Fig. \ref{f-sim-result}d. Through these simulations, we confirm that evanescent codewords are redundant for CSI feedback.

\section{Are Evanescent components present in Near-field and Rayleigh Channels?}\label{sec-5}
Near-field MIMO communications, with their potential to further enhance spectral efficiency, have been extensively studied in recent years, as summarized in \cite{cuim23, anjc24}. This section investigates whether near-field channels support the propagation of evanescent waves to provide guidance for codebook design in near-field MIMO. Additionally, the Rayleigh channel is discussed regarding the presence of evanescent components.
\subsection{Near-field Channels}\label{sec-nf}
We will restrict our discussion to the Fresnel (radiative) near-field region, typically in the range $[0.62\sqrt{\frac{D^3}{\lambda}},\frac{2D^2}{\lambda})$ \cite{balanis15}, where the EM field has negligible radial component and the propagation can be well approximated by the spherical wave model. For a UPA lying in the $x$-$o$-$y$ plane as shown in Fig. \ref{f-xyz}a, to focus the transmission energy at the UE location $(r,\theta,\varphi)$, the phase delay for the antenna locating at $(x, y, 0)$ is determined by the distance between the antenna and the UE, i.e., $\phi(x,y)=kr(x,y)$, where the distance $r(x,y)$ is calculated as $r(x,y)=\sqrt{\xi^2+\eta^2+\zeta^2}$, with $\xi=x-r\sin\theta\cos\varphi$, $\eta=y-r\sin\theta\sin\varphi$ and $\zeta=r\cos\theta$. Compared with a LOS far-field channel, a LOS near-field channel does not pertain a constant phase gradient.

\begin{proposition}\label{prop-2}
Fresnel near-field channels do not support the propagation of evanescent components.
\end{proposition}
\begin{IEEEproof}
The spatial frequency of a near-field channel is determined by the phase gradient, i.e., $k_x=\frac{\partial \phi(x,y)}{\partial x}$ and $k_y=\frac{\partial \phi(x,y)}{\partial y}$. According to the expression of $\phi(x,y)$, we obtain
\begin{equation}
k_x=\frac{k \operatorname{sgn}(\xi)}{\sqrt{1+\frac{\eta^2+\zeta^2}{\xi^2}}},\quad k_y=\frac{k \operatorname{sgn}(\eta)}{\sqrt{1+\frac{\xi^2+\zeta^2}{\eta^2}}}.\label{eq-kxky-nf}
\end{equation}
Since $\sqrt{1+\frac{\eta^2+\zeta^2}{\xi^2}}\geq 1$ and $\sqrt{1+\frac{\xi^2+\zeta^2}{\eta^2}}\geq 1$, we have $|k_x|\leq k$ and $|k_y|\leq k$. The transverse spatial frequency $|\boldsymbol{k}_t|$ can then be calculated as following, 
\begin{equation}
|\boldsymbol{k}_t|=\sqrt{k_x^2+k_y^2}
=\frac{k}{\sqrt{1+\frac{\zeta^2}{\xi^2+\eta^2}}}\leq k.
\end{equation}
Consequently, the spatial frequencies of Fresnel near-field channels are bounded by $k$. Evanescent components with spatial frequencies exceeding $k$ are not supported by these channels.
\end{IEEEproof}

\begin{figure}[t]
\centering
\includegraphics[width=0.48 \textwidth]{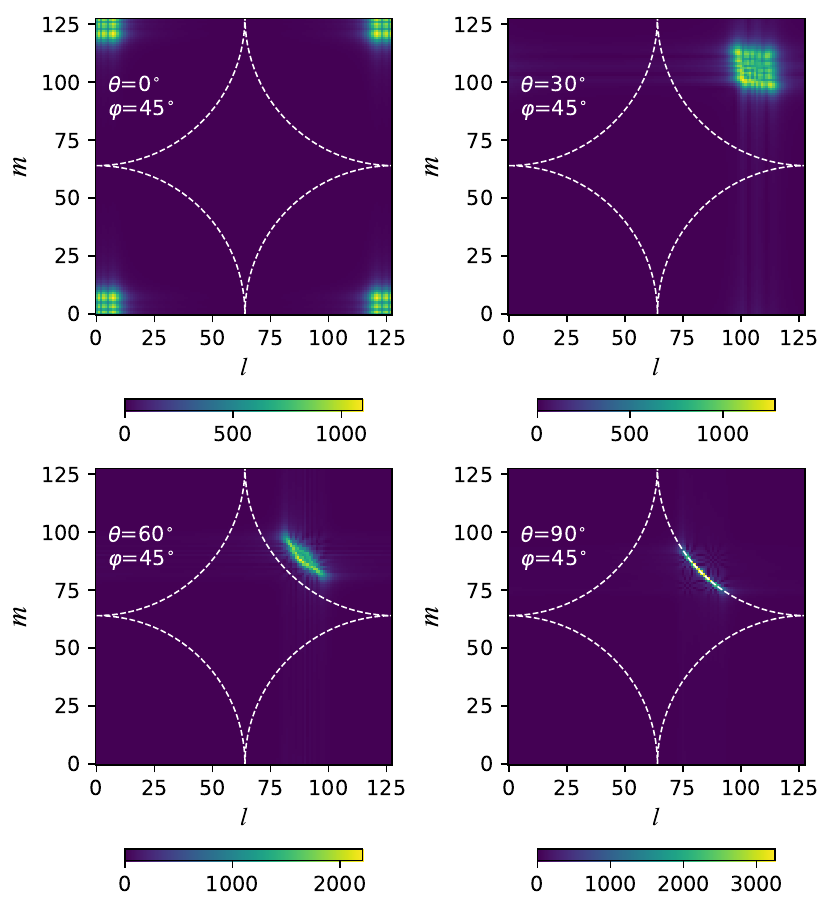}
\caption{The projections of 4 near-field channels onto $\mathbf{CB}_{128,128}^{1,1}$.}
\label{f-nf-dft}
\end{figure}

According to Proposition \ref{prop-2}, Fresnel near-field channels are also low-pass spatial filters. It will be futile to attempt to utilize evanescent waves for communications in the Fresnel near-field region. When Kronecker product-based codebooks are used for near-field channel representation, only propagating codewords are necessary, making Algorithm \ref{alg-1} applicable to Fresnel near-field MIMO systems. This can be readily verified by projecting the near-field channel onto a codebook. As shown in Fig. \ref{f-nf-dft}, four near-field channels between a UPA with $128 \times 128$ half-wavelength antennas and four single-antenna UEs were projected onto the codebook $\mathbf{CB}_{128,128}^{1,1}$. The selected UE locations were $\mathbf{r}_1 = (D, 0^{\circ}, 0^{\circ})$, $\mathbf{r}_2 = (D, 30^{\circ}, 45^{\circ})$, $\mathbf{r}_3 = (D, 60^{\circ}, 45^{\circ})$, and $\mathbf{r}_4 = (D, 90^{\circ}, 45^{\circ})$, where $D = 64\sqrt{2}\lambda$ is the maximum dimension of the UPA. As expected, the projected energy spreads but remains confined within the propagating zone for all four near-field channels. In the lower-right subfigure, the energy extends close to the zone boundary and slightly diffuses into the evanescent zone. This minor diffusion is attributed to the finite aperture of the UPA, as explained in Section \ref{sec-sys-sim}.

Clearly, if dedicated near-field codebooks are to be designed, the codewords should not contain spatial frequencies representing evanescent waves. According to \eqref{eq-kxky-nf}, the spatial frequency of a Fresnel near-field channel varies with the position of the transmitting antenna, a phenomenon known as space-frequency localization \cite{goodman05}. This localization enables the use of non-uniform arrays for near-field transmissions.

\subsection{Rayleigh Channels}
\begin{figure}[t]
\centering
\includegraphics[width=0.44 \textwidth]{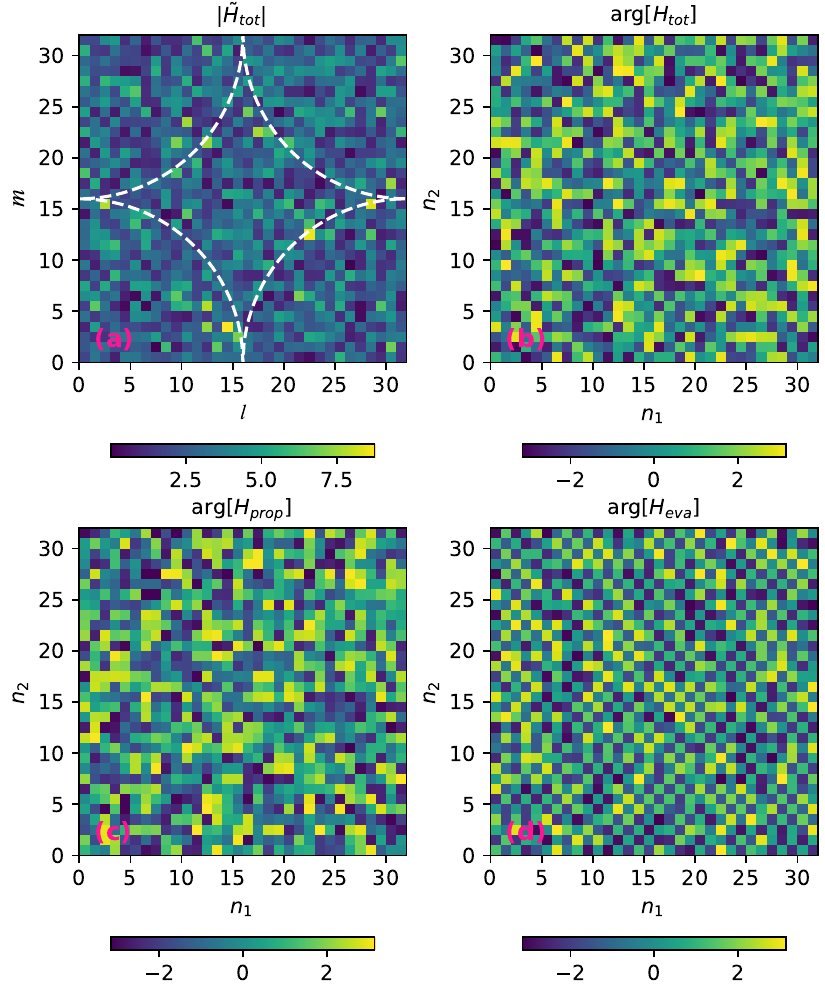}
\caption{Projection and phase distributions of a Rayleigh channel: (a) Projection of $\mathbf{H}_{\text{tot}}$ onto $\mathbf{CB}_{32,32}^{1,1}$; (b) Phase distribution of $\mathbf{H}_{\text{tot}}$; (c) Phase distribution of $\mathbf{H}_{\text{prop}}$; (d) Phase distribution of $\mathbf{H}_{\text{eva}}$.}
\label{f-rayleigh}
\vspace{-1em}
\end{figure}

The Rayleigh channel model has been widely used for evaluating the performance of various MIMO systems and algorithms due to its straightforward mathematical formulation and favorable statistical properties (e.g., \cite{bravo13,mathur14,bedeer16}). However, considering the evanescent codewords discussed in this paper, questions arise regarding whether the randomly generated Rayleigh channel accurately preserves the low-pass property of a physical channel. This concern can be addressed using the projection method described in Section \ref{sec-nf}. For clarity, let $\mathbf{H}_{\text{tot}}$ denote the generated Rayleigh channel, while $\tilde{\mathbf{H}}_{\text{tot}}$ represents its DFT, which is equivalent to a projection onto a codebook with $O_1=O_2=1$. By generating a series of Rayleigh channels and applying the projection, we observed that each of these channels contained evanescent components. Figs. \ref{f-rayleigh}a and \ref{f-rayleigh}b show the projected amplitude on $\mathbf{CB}_{32,32}^{1,1}$ and the spatial phase distribution, respectively, for one example test case. Clearly, the channel energy is distributed across both propagating and evanescent zones. The zone boundary can be identified with Proposition \ref{prop-1}, allowing us to separate the propagating component, $\mathbf{H}_{\text{prop}}$, from the evanescent component, $\mathbf{H}_{\text{eva}}$, within $\mathbf{H}_{\text{tot}}$. Figs. \ref{f-rayleigh}c and \ref{f-rayleigh}d show the phase distributions of $\mathbf{H}_{\text{prop}}$ and $\mathbf{H}_{\text{eva}}$, illustrating the low and high spatial variations of the propagating and evanescent components, respectively.

We can now conclude that a randomly generated Rayleigh channel contains both propagating and evanescent components. Since evanescent components will be filtered out by physical far-field and near-field channels, they are artifactual and cannot be utilized for transmission. Consequently, the capacity of a Rayleigh channel may be overestimated. To address this issue, we propose the following four-step method for generating Rayleigh channels:
\begin{enumerate}
    \item Generate a channel using the classical Rayleigh channel model.
    \item Perform a DFT on the randomly generated channel.
    \item Zero out the amplitudes of components in the evanescent zone.
    \item Transform the modified channel back to the spatial domain using an IDFT.
\end{enumerate}
These four steps ensure that the generated Rayleigh channel contains no evanescent components and the performance evaluation will be more reliable. For non-isotropic scattering environments \cite{pizzo22}, a more accurate Rayleigh channel model can be constructed by eliminating spatial frequency components that are absent in the wavenumber domain with our proposed four-step generation method.

\section{Conclusion}\label{sec-6}
In this paper, we establish a foundational link between array precoding and EM wave propagation, highlighting that precoding is not merely a signal processing technique but is deeply intertwined with the physical properties of EM waves. We identify the presence of evanescent codewords in Kronecker product-based codebooks and explore their mathematical and physical implications. Mathematically, these codewords correspond to imaginary beam angles, while physically, they represent evanescent EM components with high spatial frequencies that cannot be efficiently radiated. Through array synthesis theory, spatial sampling theory, and system-level simulations, we demonstrate that evanescent codewords are redundant and should be excluded from the codebooks. Based on the spatial frequency theory, we also propose Kronecker product-type codebooks for irregular arrays.

Considering the redundancy, we propose a method for compressing the codebooks. While this compression has negligible impact on the reduction of CSI feedback overhead, it significantly decreases the overhead associated with CBSR signaling and beam training pilots. Additionally, the efficiency of codeword selection in codebook-based CSI reporting and the beam training process is greatly improved without introducing additional algorithmic complexity. Finally, we show that Fresnel near-field channels do not support the propagation of evanescent EM components, offering valuable insights for future near-field codebook design. Furthermore, we propose a modified Rayleigh channel generation method to eliminate artifactual evanescent components, enhancing the practical applicability of the channel model.

\end{document}